\documentclass{article}
\usepackage[english]{babel}
\usepackage[utf8]{inputenc}
\usepackage{amsmath}
\usepackage{amsfonts}
\usepackage{amsthm}
\usepackage{authblk}
\usepackage{xcolor}
\usepackage{tikz}
\usepackage{listings}
\usepackage[margin=1in,footskip=0.25in]{geometry}
\usepackage{url}
\usepackage[ruled,vlined]{algorithm2e}
\usepackage{bbm}

\newtheorem{theorem}{Theorem}

\theoremstyle{definition}

\newtheorem{proposition}[theorem]{Proposition}

\title{Treatment effect estimation with Multilevel Regression and Poststratification}
\author[$1$]{Yuxiang Gao \thanks{Author email: \texttt{ygao@utstat.toronto.edu}}}
\author[$2$]{Lauren Kennedy \thanks{Author email: \texttt{lauren.kennedy1@monash.edu}}}
\author[$1$]{Daniel Simpson \thanks{Author email: \texttt{d.simpson@utoronto.ca}}}
\affil[$1$]{Department of Statistical Sciences, University of Toronto}
\affil[$2$]{Department of Econometrics and Business Statistics, Monash University}

\usetikzlibrary{shapes,decorations,arrows,calc,arrows.meta,fit,positioning}
\tikzset{
    -Latex,auto,node distance =1 cm and 1 cm,semithick,
    state/.style ={circle, draw, minimum width = 0.7 cm},
    point/.style = {circle, draw, inner sep=0.04cm,fill,node contents={}},
    bidirected/.style={Latex-Latex,dashed},
    el/.style = {inner sep=2pt, align=left, sloped}
}

\usetikzlibrary{bayesnet}
\usetikzlibrary{arrows}

\begin{document}

\maketitle

\begin{abstract}
    Multilevel regression and poststratification (MRP) is a flexible modeling technique that has been used in a broad range of small-area estimation problems. Traditionally, MRP studies have been focused on non-causal settings, where estimating a single population value using a nonrepresentative sample was of primary interest. In this manuscript, MRP-style estimators will be evaluated in an experimental causal inference setting. We simulate a large-scale randomized control trial with a stratified cluster sampling design, and compare traditional and nonparametric treatment effect estimation methods with MRP methodology. Using MRP-style estimators, treatment effect estimates for areas as small as 1.3$\%$ of the population have lower bias and variance than standard causal inference methods, even in the presence of treatment effect heterogeneity. The design of our simulation studies also requires us to build upon a MRP variant that allows for non-census covariates to be incorporated into poststratification.
\end{abstract}


\let\thefootnote\relax\footnote{\textbf{Acknowledgements: }Yuxiang Gao was funded by the Ontario Graduate Scholarship. Daniel Simpson was funded by the Natural Sciences and Engineering Research Council of Canada and the Canadian Research Chair program. Research reported in this publication was supported by National Institute of Aging of the National Institutes of Health under award number R01AG067149. The content is solely the responsibility of the authors and does not necessarily represent the official views of the National Institutes of Health. }

\section{Introduction}

Randomized control trials (RCTs) are often considered the gold standard for estimating causal treatment effects of a specific intervention. Applications of RCTs are widespread, having been applied in fields such as healthcare, economics, political science, education and technology. Studying the statistical properties of RCTs and the methods used to model them are active areas of research in causal inference since challenges involving data quality, model specification and experimental design can arise in practice.

The main purpose of this manuscript is to introduce Multilevel Regression and Poststratification (MRP) \cite{gelman1997poststratification, little1993post}, a method has traditionally been used for public opinion modelling and small-area estimation, as a viable method in experimental causal inference. We will highlight the strengths and limitations of modern causal inference methods along with MRP and its variants for estimating conditional treatment effects over subpopulation groups of various sizes, focusing specifically on scenarios with treatment effect heterogeneity. In particular, the causal inference problem we are analyzing is a simulated version of a large-scale stratified cluster RCT conducted in the United States in $2015$--$2016$. Secondly, we analyze a variant of MRP seen in \cite{lax2019party, mrpnoncensus, kastellec2015polarizing}, which we call \textit{Multilevel Regression and Poststratification with Multiple Imputation (MRP-MI)}. MRP-MI allows  the modeler to include a variable that is present in the sample but missing in the population, known as a non-census variable, in the poststratification frame.

\subsection{The notion of generalization for causal estimates}

\textit{Generalization} in causal inference refers to mapping treatment effects from a specific study to treatment effects in a target population. This population can be a finite population of the experimental sample, a group of individuals who are in a different location relative to the experimental sample or a group of individuals whose covariates were measured at a different time relative to the experimental sample. Without post-experimentation adjustments, there is a chance that treatment effects from an experimental sample will be different than the true treatment effect of the target population. An important point to note is that, no unadjusted treatment effect estimates from the experimental sample are generalizable to the target population unless the sample is equal to the target population or the sample is a random draw from the target population \cite{stuart2018generalizability}. There are different interpretations and definitions of generalizability in the literature, and in this part of the introduction we will briefly outline some of the different frameworks and assumptions used to define generalizability. 

In \cite{kern2016assessing}, the estimand of interest is the Target Average Treatment Effect (TATE), which is defined as the mean of the difference in potential outcomes $\frac{1}{N}\sum_{i=1}^N\left(Y_i(1) - Y_i(0) \right)$ where $N$ is the size of the target population. A sampling indicator $S_i \in \{0,1\}$ is defined for every individual in the target population and the experimental sample. $S_i=0$ if the individual is in the experimental sample and $1$ otherwise. Estimating TATE can be done through weighting approach, or an outcome modelling approach. Through the weighting approach as outlined in \cite{kern2016assessing}, a score $w_i:=\frac{\hat{e}_i}{1-\hat{e}_i}$ is defined for every individual in the experimental sample, where $\hat{e}_i$ is the estimated probability that individual $i$ is in the target population. The idea behind the weights $w_i$ is that they make the experimental sample more similar to the target population. A weighted linear regression model is then fit to the experimental sample using the set of $w_i$ and then the coefficient of the treatment indicator is used as the TATE estimate. The outcome modelling approach in \cite{kern2016assessing} uses a predictive model on the experimental sample and then predicts outcomes under treatment and control for every individual in the the target population.

In \cite{ackerman2019implementing}, the generalization methods of weighting and outcome modelling as seen in \cite{kern2016assessing} are tested.  Targeted Maximum likelihood Estimation \cite{gruber2009targeted}, a doubly robust estimation method that combines both weighting and outcome modelling is used to estimate the TATE. For a more thorough literature review of generalizing experimental sample treatment effect estimates to the TATE using approaches in \cite{ackerman2019implementing,kern2016assessing}, the reader can refer to \cite{stuart2018generalizability}. A broader literature review on experimental causal inference and generalizability can be found in the recent survey \cite{colnet2020causal}.%

Another way to generalize experimental sample estimates to a target population is through poststratification, a weighting technique in survey statistics. Poststratification is briefly mentioned in \cite{stuart2011use}, but the current challenges that remain are continuous covariates or a large number of poststratification cells. \cite{miratrix2018worth} analyzes poststratification in survey experiments, but only analyzes scenerios where all observed covariates are categorical. 

To measure representativeness between the experimental sample and the target population, various measures have been developed. One method was based off the \textit{sampling propensity score} $s(X):=\mathbb{P} (S=1|X)$ where $X$ is the pre-treatment covariate vector and $S$ is a binary indicator for membership in the experimental sample. \cite{stuart2011use} define the propensity score difference, which is the difference of averages for $s(X_i)$ between the experimental sample and the target population. \cite{tipton2014generalizable} extends the usage of sampling propensity scores by defining a new metric known as the \textit{generalization index}, a value between 0 and 1 that is a measure of representativeness between the experimental sample and the target population. Though this manuscript is not focused on generalizing treatment effect estimates but rather on recovering treatment effect heterogeneity, we calculate the generalization index in our simulation studies to show that indeed our experimental sample is representative and hence a high-quality sample of the target population.

\subsection{Treatment effect heterogeneity across different subpopulations in the target population}

When there is no treatment effect heterogeneity, it's not a concern if the experimental sample is representative or not. This is because treatment effects for any subpopulation will be equal so over/undersampling certain subpopulations does not result in different treatment effect estimates. However, any amount of treatment effect heterogeneity will bias treatment effect estimates coming from a nonrepresentative experimental sample when the nonrepresentativeness and heterogeneity are not properly adjusted for. 

More specifically, estimating heterogeneous treatment effects consists of estimating the non-constant function $\tau(x):=\mathbb{E} \left(Y(1) - Y(0) | X= x \right)$, which is the expected value of the difference in potential outcomes conditional on pretreatment variables $X=x$. This expectation is taken at the superpopulation level, where the superpopulation is the data generating process for the finite target population. In this manuscript, we generate a finite target population then sample from it, but the observed sample is used to estimate CATEs at the superpopulation level. For additional details on the connection between superpopulation causal inference and finite population causal inference, we refer the reader to \cite{ding2017bridging, imbens2015causal}.

Modeling $\tau(x)$ can be done parametrically and non-parametrically. The parametric approach would be to use a linear regression model with the treatment variable interacting with other pre-treament variables. A parametric modeling approach's benefits are that it's simple to implement and its model coefficients are interpretable. However, it's prone to model misspecification when $\tau(x)$ is nonlinear and pre-treatment covariate $X$ is high-dimensional, which can result in biasing estimates of $\tau(x)$.

Estimating $\tau(x)$ can be done more flexibly with nonparametric regression methods \cite{hahn2020bayesian,wager2018estimation,hill2011bayesian} and  machine learning methods \cite{kunzel2019metalearners}. Nonparametric treatment effect estimation methods have been used in survey experiments \cite{green2012modeling} and large-scale randomized control trials concerning education interventions \cite{yeager2019national,athey2019estimating}. For a more extensive overview of the challenges arising from treatment effect heterogeneity in experimentation, we refer the reader to \cite{athey2017econometrics}.

The focus of this paper will mainly be to address the challenges of treatment effect heterogeneity in experimentation when the sample is representative of the target population, so as a motivating example, we will discuss a large-scale RCT conducted recently on the target population of public high schools in the United States, one that was representative of the target population yet exhibited treatment effect heterogeneity. This RCT was used as a model for our simulation studies run in this manuscript.

\subsection{The National Study of Learning Mindsets: A large-scale randomized control trial}

The National Study of Learning Mindsets (NSLM) \cite{yeager2019national} was an experiment conducted on a nationally representative sample of $9^{\text{th}}$ grade public high-school students in the United States during Fall 2015. The \textit{growth mindset intervention} was an education intervention that encouraged treated individuals to view intellectual ability not as a fixed trait, but a muscle that can be trained through the sustained effort of seeking help when learning new academic content and trying new learning strategies. Indeed, the growth mindset intervention was shown to be effective in improving the GPA of high-school students. The original NSLM analysis used linear modeling with survey weights as well as a response surface method \cite{hahn2020bayesian}.

The sampling design in NSLM first has school strata defined by levels of minority composition level and school-achievement level. Then schools are cluster sampled from each school strata. Finally, in the chosen schools that decided to partake in the study, students were given the option to participate or not. The average student-level response rate was 92 percent. The intervention was randomly assigned at the student level.

This manuscript's simulation study is interested in determining how effective an education intervention is for the GPA of individuals. This is measured as the difference in potential outcomes of post-intervention GPA. The causal relationship of interest in this manuscript is shown in the causal directed acyclic graph (DAG) \cite{pearl1995causal} seen in Figure \ref{fig:dag}. This DAG is a simplified model of the NSLM that contains only the main variables in their analysis \cite{yeager2019national}.

$\newline$

\begin{minipage}{.48\textwidth}
  \textbf{Figure \ref{fig:dag} (Causal DAG):} All the shaded nodes are observed variables. Because the data collected is coming from a randomized experiment, all the variables except the outcome (blue post-intervention GPA node) and the treatment (red treatment node) are pre-treatment variables. The structure of the posited causal graph is incorporated in treatment effect estimation methods used in this manuscript. The blue outcome node, Post-GPA, and the grey covariate node, Prev-GPA (Previous GPA of an individual), are both continuous and truncated with a lower and upper bound. The rest of the observed covariates are categorical.
\end{minipage} \hspace{\fill}
\begin{minipage}{.48\textwidth}
\resizebox{1\textwidth}{!}{

  \begin{tikzpicture}[
    ->,
    >=stealth',
    auto,node distance=3cm,
    thick,
    main node/.style={circle, draw, font=\sffamily\Large\bfseries}]
  \node[obs] (z) [color=red!60, fill=red!15, ultra thick] at (0,0) {$\textcolor{red}{\textbf{Treatment}}$};
  \node[obs] (y) [left=2cm of z, color=blue!60, fill=blue!15, ultra thick] {$\textcolor{blue}{\textbf{Post-GPA}}$};
  \node[obs] (x) [left=of y, ultra thick] {$\text{Prev-GPA}$};
  \node[obs] (s) [below=of y, ultra thick] {$\text{School}$};
  \node [obs] (re) [above= 2.5 cm of x, ultra thick] {$\text{Race/Eth}$};
  \node [obs] (me) [right=of re, ultra thick] {$\text{Mat.-Edu.}$};
  \node [obs] (g) [left=of re, ultra thick] {$\text{Gender}$};
  \node[obs] (mc) [below left =0.5 cm and 2 cm of x, ultra thick] {$\text{Min-Comp.}$};
  \node[obs](sa) [above=of mc, ultra thick] {$\text{School-Ach.}$};

     \edge{x}{y};
     \edge{s}{y};

         \edge {re,me,g} {y};

         \edge {z} {y};
         
         \edge {sa} {re};
         \edge {sa} {me};
         \edge {mc} {me};
         \edge {mc} {re};

         \draw [->] (sa) to [bend left=15] (y);
         \draw [->] (mc) to [bend right=10] (y);
         
         \draw [->] (sa) to (x);

    \label{fig:dag}     
    \end{tikzpicture}
     }

\end{minipage}

$\newline$

The NSLM's target population covariate structure is rich and detailed. Across different covariate subpopulations, different treatment effects are suspected to exist. In this manuscript, we explore the limitations of using a representative sample to calculate treatment effects for subpopulations in a population as structured and heterogeneous as the target population of NSLM.

We design a simulation study that generates a complex target population and samples from it through a stratified cluster design to generate a RCT. The sampling design has schools as the clusters, and the clusters are stratified by school-level covariates minority composition level and school-achievement level. The simulated population is built to match the population summary statistics reported in \cite{yeager2019national}. Section 5 contains additional details on the design of the simulation study and further information about the population's structure and how it relates to the NSLM can be found in Supplementary Material C.

Additionally, we compare various causal inference methods to see the extent that these methods can recover the true heterogeneity of the intervention in the target population. In our simulation studies, we estimate treatment effects ranging from the Average Treatment Effect (ATE) for the whole population to Conditional Average Treatment Effects (CATEs) for subpopulations as small as 1.3 percent of the population.

\subsection{Introduction of our proposed treatment effect estimation method}

In this manuscript, we propose a treatment effect estimation method which builds on a hierarchical model-based survey estimation method used for small-area estimation, Multilevel Regression and Poststratification (MRP) \cite{park2004bayesian,gelman1997poststratification}. MRP combines a hierarchical model's posterior distribution of the outcome $(\theta_c)_{c=1}^J$ with an external poststratification matrix containing population sizes $(N_c)_{c=1}^J$, where $c$ is a stratum of finest granularity in the target population. MRP then forms a posterior distribution for the average outcome in the target population, $\theta_{ps}:=\frac{\sum_{c=1}^J N_c \theta_c}{\sum_{c'=1}^J N_{c'}}$, and summary statistics such as the posterior mean of $\theta_{ps}$ can be used as the point estimate for average outcome in the target population. A strength of MRP is that it's effective in estimating outcomes of small areas in the presence of selection bias across various subpopulations of a target population \cite{gao2020improving}. MRP has historically been a commonly used methodology in small-area estimation and public opinion modeling \cite{lauderdale2020model, wang2015forecasting}, however it's applications to causal inference have been an unexplored area. A primary goal of this manuscript is to explore the extent of MRP's effectiveness in experimental causal inference. For a more detailed step-by-step outline of MRP, we refer the reader to \cite{lax2009should}.

MRP requires covariates to be categorical, in order to perform \textit{Poststratification}. Current practice is to discretize continuous covariates before using them in a MRP model. The effect of the  granularity of the discretization on the MRP estimates is studied  in \cite{gao2020improving}. Discretization of continuous covariates in MRP is possible when we have population counts for each of the discretized bins. For example, if age is discretized into 6 groups, then the poststratification matrix used must have population counts for each of the 6 age groups. 

In our simulation study, which is based on the NSLM study, we are given the poststratification matrix for all combinations of School-Achievement $\times$ Minority-Composition $\times$ School $\times$ Gender $\times$ Race/Ethnicity $\times$ Maternal-Education, thus providing population counts for each combination. Prev-GPA of individuals in every collected sample is reported, but no such discretization of Prev-GPA at the population level is available. 

It's also important to note that Post-GPA and Prev-GPA in our simulation studies are both truncated continuous distributions with support $(0, 4.33)$. If Post-GPA and Prev-GPA were  untruncated normal distributions, then treatment effects (ATE and CATEs) in the DAG Figure \ref{fig:dag} can be estimated by modeling the untruncated normal distribution for $\text{Post-GPA} - \text{Prev-GPA}$ as a function of the binary treatment indicator and the covariates School-Achievement $\times$ Minority-Composition $\times$ School $\times$ Gender $\times$ Race/Ethnicity $\times$ Maternal-Education. This is not the case in our study since the variable $\text{Post-GPA} - \text{Prev-GPA}$ is not in the same family of distributions that Post-GPA or Prev-GPA are in. Thus to estimate the treatment effects from DAG Figure \ref{fig:dag}, we have to use Prev-GPA as a covariate and Post-GPA as the outcome.

We will utilize a variant of MRP that bypasses missing population counts for the continuous covariate Prev-GPA and also takes into account the truncated nature of both Prev-GPA and Post-GPA. We will refer to this variant as \textit{Multilevel Regression and Poststratification with Multiple Imputation (MRP-MI)}. We apply MRP-MI to to the problem of heterogeneous treatment effect estimation for the causal graph in Figure \ref{fig:dag}. 

In the literature, MRP-MI is an extension of the procedure where the modeler uses MRP with non-census variables \cite{kennedy2019know,lax2019party,mrpnoncensus, kastellec2015polarizing}. MRP-MI is similar to the prediction framework carried out in example 2 of \cite{kennedy2019know}.   \cite{kennedy2019know} dealt with a non-census variable by imputing it for each individual.
In contrast, we predict our non-census variable at the poststratification cell-level rather than the individual level.
 In \cite{lax2019party,kastellec2015polarizing} the non-census variable (variable with missing population counts) is political party membership (discrete with 3 levels) whereas in this manuscript the non-census variable is Prev-GPA (continuous and truncated).

In this estimation problem, MRP-MI entails fitting both a Prev-GPA hierarchical model and a Post-GPA hierarchical model, and then poststratifying with the posterior predictive distribution of Post-GPA after integrating out the posterior predictive distribution of Prev-GPA. The detailed description of MRP-MI is shown in the later Section 4.

The challenge with modeling CATEs in a simulation study modelled after the target population in NSLM is that there are scenarios where some school clusters $\tilde{\mathcal{O}}$ are not observed. Hence calculating the CATE for $\tilde{\mathcal{O}}$ will be an out-of-sample prediction problem. Indeed, that was the case in the original NSLM study -- Only around 70 schools out of the 12000 total schools in the target population were sampled from. Instead of calculating CATEs for all 12000 schools, the modelers only calculated CATEs for the clusters observed. A part of this manuscript (Section 5.6) will be dedicated to showcasing the limitations of MRP-MI and other MRP variants when it comes to calculating out-of-sample CATEs.

\subsection{Structure of the paper}

This manuscript has the following order: Section 1 contains a summary of two common problems in experimental causal inference, namely treatment effect heterogeneity and generalizability of treatment effect estimates. Section 2 contains our problem setup for estimating treatment effects for any group in the target population, where our problem is modeled after the National Study of Learning Mindsets. Section 3 presents the various treatment effect estimation methods that are tested in our simulation studies. Section 4 presents the detailed outline of our proposed treatment effect estimation method. Section 5 is a description of our simulation study and results, which includes how the target population is generated and how we sample from the target population. Section 6 is the conclusion. The supplementary material contains further discussion of the MRP variant (MRP-MI), connection between treatment effect estimation and MRP, and simulation design information. %

\section{Problem setup}

Suppose that we're interested in estimating how effective an education intervention is on the Post-GPA of an individual. Let $\left(Y(0), Y(1), X \right)$ be a superpopulation joint distribution for the control potential outcome, treatment potential outcome and the pre-treatment covariate. $Y(0)$ is the Post-GPA of the individual under no treatment and $Y(1)$ is the Post-GPA of the individual under treatment. $X$ is a vector containing demographic information and the previous academic achievement scores of an individual. We're interested in the following superpopulation CATE estimands:

\begin{enumerate}
    \item Average Treatment Effect: $\tau_\text{ATE}:=\mathbb{E} \left( Y(1) - Y(0) \right)$
    \item Conditional Average Treatment Effect for any subpopulation $\mathcal{O}$: $\tau_{\text{CATE},\mathcal{O}}:=\mathbb{E}\left(Y(1) - Y(0) | X \in \mathcal{O} \right)$
    \item Conditional Average Treatment Effect for $X=x$: $\tau_{\text{CATE},X=x}:=\mathbb{E}\left(Y(1) - Y(0) | X = x \right)$
\end{enumerate}

Suppose that a finite population $P:=\left(Y_m(0),Y_m(1), X_m \right)_{m=1}^{N_P}$ is generated $i.i.d$ from $\left(Y(0), Y(1), X \right)$. Suppose that $\left(Y_i(0),Y_i(1), X_i \right)_{i=1}^{n}$ is a sample drawn from $P$ in a probabilistic manner. In our case, $\left(Y_i(0),Y_i(1), X_i \right)_{i=1}^{n}$ is drawn as a stratified cluster sample. Let $Z_i \in \{0,1\}$ be a treatment-control indicator variable for individual $i$ in the sample, and $Y_i$ be the outcome variable for individual $i$ in the sample. The details of the sampling design and assignment mechanism for $\left(Y_i(0),Y_i(1), X_i, Z_i \right)_{i=1}^{n}$ can be found in Section 5 of this manuscript.

Referencing the causal DAG in Figure \ref{fig:dag}, we have $X_i = \left( V_i,\text{ME}_i,\text{G}_i,\text{RE}_i, \text{School}_i,\text{SA}_i,\text{MC}_i 
\right)$, which is the pre-treatment covariate vector for individual $i$ consisting of Previous-GPA, Minority Composition, Gender, Race/Ethnicity, School, School Achievement level and Minority Composition level of School. School Achievement level and Minority Composition level of School are school-level covariates and the other covariates are individual-level covariates. $\left(Y_i(0),Y_i(1), X_i, Z_i \right)_{i=1}^{n}$ is a stratified cluster sample of the finite population $P$, where clusters are schools. 

Finally, we will assume that we're given a poststratification matrix $M$ for the target population, with population counts for all $J$ combinations of School-Achievement $\times$ Minority-Composition $\times$ School $\times$ Gender $\times$ Race/Ethnicity $\times$ Maternal-Education. Table \ref{ps_mat} shows $M$. We do not have population information for Prev-GPA.

\begin{table}[t]
\centering
\resizebox{\columnwidth}{!}{
    \begin{tabular}{rrrrrrr}
      \hline
Maternal Education & Gender & Race/Ethnicity & School & Minority Composition Index & School Achievement Index & $N_c$ \\ 
      \hline
No & Female &   Asian & 00001 & Low &   Low &   1 \\ 
No & Female &   Asian & 00002 & Low &   Low &   1 \\ 
No & Female &   Asian & 00006 & Low &   Low &   1 \\ 
No & Female &   Asian & 00007 & Low &   Low &   1 \\ 
No & Female &   Asian & 00012 & Low &   Low &   2 \\ 

\vdots & \vdots &   \vdots & \vdots & \vdots &   \vdots &   \vdots \\ 
\vdots & \vdots &   \vdots & \vdots & \vdots &   \vdots &   \vdots \\

Yes & Male &   Other & 11220 & High &   High &   9 \\ 
Yes & Male &   Other & 11221 & High &   High &  15 \\ 
       \hline
    \end{tabular}
}
\caption{Poststratification matrix $M$ for the population $P$ with population sizes $N_c$ for all $J$ combinations of School-Achievement $\times$ Minority-Composition $\times$ School $\times$ Gender $\times$ Race/Ethnicity $\times$ Maternal-Education. Note that $\sum_{c=1}^J N_c = N_P$.}
\label{ps_mat}
\end{table}

\subsection{Identifiability assumptions for target estimands}

To allow for identifiability of superpopulation CATEs and ATE, we will assume the following below for all individuals $1 \le i \le n$. For more on the role that these assumptions play in superpopulation treatment effect estimation, we refer the reader to \cite{imbens2015causal}.

\begin{enumerate}
    \item Stable Unit Treatment Value Assumption (SUTVA) \cite{rubin1980randomization}: We assume that there is no interference between units. That is, the treatments assigned to a unit do not affect the potential outcomes of another unit.
    \item Consistency: $Y_i = Z_i Y_i(1) + (1-Z_i) Y_i(0)$ almost-surely
    \item Ignorability: $(Y_i(0),Y_i(1)) \perp\!\!\!\perp Z_i$
    \item Overlap: $ 0 < \mathbb{P} \left( Z_i = 1 | X_i \right) < 1$ almost-surely
\end{enumerate}

From the consistency assumption, we have $Y_i = Y_i(Z_i)$ almost-surely. This defines the \textit{observed} stratified cluster sample $\mathcal{D}:=(Y_i, X_i, Z_i)_{i=1}^n$, consisting of the outcome, pre-treatment covariate and treatment variable respectively. Based off $\mathcal{D}$, various causal inference methods can be use to estimate the ATE and CATEs.

MRP-style estimators will be compared with other standard causal inference methods for calculating CATEs and ATE, so we motivate the need for MRP by providing the connection between MRP and superpopulation treatment effect estimation below.

\subsection{Connection between MRP and treatment effect estimation}

Suppose that the covariate vector $X$ of an individual in the superpopulation has $J$ number of possible values that it can take on. Also, suppose we're given a poststratification matrix $\tilde{M}$ for the target population, with population counts $(N_c)_{c=1}^J$ for all $J$ cells. Note that the CATE $\tau_{\text{CATE},\mathcal{O}}$ for a subpopulation $\mathcal{O}$ is then equal to

\begin{equation}
\begin{aligned}
    \mathbb{E}\left(Y(1) -Y(0) | X \in \mathcal{O} \right) &=\int_{c \in I_\mathcal{O}} \mathbb{E}\left(Y(1)-Y(0) | X = x_c \right) \mathbb{P}\left(X = x_c | X \in \mathcal{O} \right)\\
    & = \sum_{c \in I_\mathcal{O}} \mathbb{E}\left( Y(1)-Y(0) | X = x_c \right) \frac{N_c}{\sum_{c' \in I_\mathcal{O}}N_{c'}}\\
\end{aligned}
\end{equation}

where $I_\mathcal{O}$ is the index set for subpopulation $\mathcal{O}$. Thus poststratifying the CATE of every cell in subpopulation $\mathcal{O}$ is performing integration of the potential outcome surface difference over the covariate space, where the discrete probability measure is defined by the given poststratification weights $\left\{ \frac{N_c}{\sum_{c' \in I_\mathcal{O}}N_{c'}} \right\}_{c \in I_\mathcal{O}}$. 

It follows that the ATE is integrating $\mathbb{E}\left(Y(1)-Y(0) | X = x_c \right)$ over all the poststratification cells:

\begin{equation}
\begin{aligned}
    \text{ATE} &= \mathbb{E}\left(Y(1) -Y(0) \right)\\
    &=\int_{c } \mathbb{E}\left(Y(1)-Y(0) | X = x_c \right) \mathbb{P}\left(X = x_c  \right)\\
    & = \sum_{c=1}^J \mathbb{E}\left( Y(1)-Y(0) | X = x_c \right) \frac{N_c}{\sum_{c'=1}^JN_{c'}}\\
\end{aligned}
\end{equation}

Hence calculating the superpopulation ATE and superpopulation CATEs require modelling the response surfaces $g_z(x) := \mathbb{E} \left(Y(z) | X = x \right)$ for $z=0,1$, and then poststratifying the difference $\Delta g (x) : = g_1(x) - g_0(x)$ by weights defined by $\left\{ \mathbb{P} \left( X = x_c \right) \right\}_{c=1}^J$ and $\left\{ \mathbb{P}\left( X = x_c | X \in \mathcal{O} \right) \right\}_{c \in I_\mathcal{O}}$ respectively. Various outcome modeling methods as mentioned in the introduction can be used to model the two potential outcome surfaces $g_0(x), g_1(x)$, with an increasingly popular approach being nonparametric methods. Nonparametric methods such as BART have performed well in empirical studies for estimating treatment effects in an observational setting such as \cite{hahn2020bayesian,dorie2019automated}.

Poststratification weights and sampling weights (Both are survey weights) are used to adjust for selection bias of the covariates that are not the treatment covariate. On the other hand, propensity score weights in causal inference are used to adjust for non-random treatment assignment. Both can be combined together when modeling to simultaneously adjust for selection bias across non-treatment covariates and reduce confounder bias in treatment effect estimates \cite{dugoff2014generalizing}. Weights in survey statistics and weights in causal inference are related insofar that they both adjust for representativeness---survey weights adjust for representativeness in the sample whereas causal inference weights adjust for representativeness between treatment and control groups. This connection is discussed in detail in \cite{mercer2017theory}.

Another connection between survey statistics and causal inference is that both fields rely on the same machinery from missing data analysis as ways of accounting for nonresponse and confounders respectively. In Rubin's potential outcomes framework \cite{rubin1974estimating}, the target estimand is a treatment effect in the population, which is defined as a difference in potential outcomes. In survey statistics, the target estimand is a population-level quantity. Both fields require the modeler to predict  missing values in the target population. In causal inference, every individual in the observed sample has a potential outcome missing that must be predicted. In survey statistics, the outcome (and potentially some covariates) must be predicted for individuals that did not respond. \cite{kang2007demystifying} provides a more detailed review of methods that can adjust for missing data in both survey estimation problems and in causal inference problems. This paper was inspired by the commonality of both fields using the same methods as seen in \cite{kang2007demystifying} to adjust for missing data. MRP has historically been a method used for nonresponse adjustment in surveys, and in this paper we explore its capabilities in experimental causal inference.

In a MRP application setting, we're given the poststratification weights $\left\{ \frac{N_c}{\sum_{c' \in I_\mathcal{O}}N_{c'}} \right\}_{c \in I_\mathcal{O}}$ for subpopulation $\mathcal{O}$. In scenarios where the modeler wants to estimate the CATE for subpopulation $\mathcal{O}$ but isn't given the poststatification weights, the in-sample frequencies of every cell $c$ in subpopulation $\mathcal{O}$ is used instead \cite{hahn2020bayesian,dorie2019automated,hill2011bayesian}. That is, $\mathbb{P}\left(X = x_c | X \in \mathcal{O} \right)= \frac{N_c}{\sum_{c' \in I_\mathcal{O}}N_{c'}}$ is replaced with $\frac{n_c}{\sum_{c' \in I_\mathcal{O}}n_{c'}}$, where $n_c$ are the number of observations in our observed sample $\mathcal{D} = \left( Y_i, X_i, Z_i \right)_{i=1}^n$ that are from poststratification cell $c$. 

Thus the point estimate of $\tau_{\text{CATE},\mathcal{O}}$ becomes 

\begin{equation}
    \begin{aligned}
        \sum_{c \in I_\mathcal{O}} \underbrace{\mathbb{\hat{E}}\left(Y(1)-Y(0) | X = x_c \right)}_\text{Point estimate from a model} \overbrace{\mathbb{\hat{P}}\left(X = x_c | X \in \mathcal{O} \right)}^\text{frequency from $\mathcal{D}$} &=\sum_{c \in I_\mathcal{O}} \underbrace{\mathbb{\hat{E}}\left(Y(1)-Y(0) | X = x_c \right)}_\text{Point estimate from a model} \left( \frac{n_c}{\sum_{c' \in I_\mathcal{O}}n_{c'}} \right)
    \end{aligned}
    \label{eqn:cateOinsample}
\end{equation}

This can result in a poor point estimate of $\tau_{\text{CATE},\mathcal{O}}$ if the covariates in the observed sample $\{ X_i \}_{i=1}^n$ are a nonrepresentative sample of the target population. Supplementary Material B further expands on such challenges.

Assuming that a poststratification matrix $\tilde{M}$ is available, it would  be sensible to perform poststratification with it in the scenario of treatment effect estimation. Doing so will help the modeler adjust for non-response and sampling bias.

\section{Estimation methods for heterogeneous treatment effects in a simulation study modeled after NSLM}

We propose five treatment estimation methods below. The first three treatment methods (OLS, SVY, BART) had variants of them used in the original NSLM study \cite{yeager2019national} for ATE and CATE estimation of various subpopulations, hence we decided to include them in our simulation study that modeled the NSLM's target population and experimental design. The last two methods (BARP-I and MRP-I) are MRP-style methods that incorporate information from poststratification matrix $M$. BARP-I and MRP-I impute a Prev-GPA estimate for every poststratification cell in $M$, hence the acronym I.

All of the five methods are able to estimate the ATE $\tau_\text{ATE}$ and any CATE $\tau_{\text{CATE},\mathcal{O}}$ where subpopulation $\mathcal{O}$ has atleast one observation in the sample $\mathcal{D}$. The last two methods (BARP-I and MRP-I) are able to estimate CATEs for subpopulations not observed in $\mathcal{D}$.  BARP-I and MRP-I should be viewed more as frameworks rather than specific methods since they allow for various ways to model the outcome Post-GPA and the covariate Prev-GPA.

\begin{enumerate}
\item \textbf{Ordinary least squares (OLS)}

In this manuscript, we will follow notation consistent with \cite{gelman2006data}. For the $i^{th}$ individual, define $\beta_{j[i]}^\text{RE}$ to be their unique race/ethnicity coefficient. As well for the $i^{th}$ individual, define $\beta_{j[i]}^\text{School}$ to be their unique school coefficient. Fit the following regression model below:

\begin{equation}
\begin{aligned}
Y_i &\sim \mathcal{N}\left(\beta_0 + \beta_Z Z_i + \text{Offset}(V_i) + \beta_{j[i]}^\text{School} + \beta_{j[i]}^\text{RE} + \beta_{ME} \text{ME}_i + \beta_G \text{G}_i, \sigma^2 \right)\\
\end{aligned}
\label{eqn:svyreg_model}
\end{equation}

Using the full sample $\mathcal{D}$, the ATE estimate would be $\hat{\beta_Z}$. Calculating CATE for a specific subpopulation in the population is done by subsetting the stratified cluster sample $\mathcal{D}$ into the targeted subpopulation and then estimating $\hat{\beta_Z}$.

\item \textbf{Survey regression with sampling weights defined through raking (SVY) \cite{lumley2020}}

Define raked weights \cite{lohr2009sampling,deming1940least} on the stratified cluster sample using the raking variables strata, gender and race/ethnicity. The population counts used for raking are defined in the poststratification matrix of $J$ cells. Using these raked weights $w_i$, fit the model for the stratified cluster sample:

\begin{equation}
\begin{aligned}
Y_i &\sim \mathcal{N}\left(\beta_0 + \beta_Z Z_i + \text{Offset}(V_i) + \beta_{j[i]}^\text{RE} + \beta_{ME} \text{ME}_i + \beta_G \text{G}_i, \frac{\sigma^2}{w_i} \right)\\
\end{aligned}
\label{eqn:svyreg_model}
\end{equation}

The weights are rescaled to sum to $1$ for numerical stability. We will use the \texttt{survey} package \cite{lumley2020} in \texttt{R} to fit this model. Note that there is no school factor since the weights $w_i$ already account for the cluster sampling within every school stratum. 

Using the full sample $\mathcal{D}$, the ATE estimate would be $\hat{\beta_Z}$. Calculating CATE for a specific subpopulation in the population is done by subsetting the stratified cluster sample $\mathcal{D}$ into the targeted subpopulation and then estimating $\hat{\beta_Z}$. The model-robust weighted standard error \cite{lumley2017fitting} for $\hat{\beta_Z}$ is computed using the \texttt{survey} package in \texttt{R} \cite{lumley2020}.

\item \textbf{Bayesian Additive Regression Trees (BART) \cite{hill2011bayesian,chipman2010bart}}

Referencing Figure \ref{fig:dag}, we have pre-treament covariate $X_i = \left( V_i,\text{ME}_i,\text{G}_i,\text{RE}_i, \text{School}_i,\text{SA}_i,\text{MC}_i 
\right)$.

\begin{enumerate}
    \item Fit the following Bayesian Additive Regression Tree (BART) model to the stratified cluster sample $\mathcal{D}$: 
    
        \begin{equation}
        \begin{aligned}
        Y_i | \{M_j \}_{j=1}^m, \{T_j \}_{j=1}^m, \epsilon_i, \sigma &\sim \sum_{j=1}^mg_j\left(Z_i, X_i; T_j, M_j \right) + \epsilon_i\\
        M_j|T_j & \overset{\text{ind.}}{\sim} \mathcal{P}_\text{Node}\\
        T_j &\overset{\text{ind.}}{\sim} \mathcal{P}_\text{Tree} \\
        \epsilon_i | \sigma^2 &\overset{\text{i.i.d}}{\sim} \mathcal{N}(0,\sigma^2) \\
        \sigma^2 &\sim \mathcal{P}
        \end{aligned}
        \label{eqn:bart}
        \end{equation}
        
        where $g_j$ is the assignment function for tree $j$, $T_j$ is tree $j$ where prior probabililty is assigned to a node splitting at each depth, $M_j:=\{ u_{jb}\}_{b=1}^{B_j}$ are the leaf nodes for tree $j$ with gaussian priors assigned to each leaf $u_{jb}$. Each independent tree has prior mass that favors shorter trees and leaf node values near $0$. There is a conjugate prior used for $\mathcal{P}$, which is usually a inverse chi-squared with data-informed hyperparameters. For additional information on the BART model specification, refer to \cite{chipman2010bart}. The \texttt{R} package \texttt{dbarts} \cite{dorie2018dbarts,chipman2010bart} is used for BART fitting.
        
    \item 
    
    The model $(\ref{eqn:bart})$ is fit to the stratified cluster sample. The posterior distribution for individual $i$ in the stratified cluster sample is defined as

    \begin{equation}
        \begin{aligned}
            \tau^{\text{BART}} \left( X_i\right) &:= \sum_{j=1}^mg_j^\text{post}\left(Z_i = 1, X_i; T_j, M_j \right) - \sum_{j=1}^mg_j^\text{post}\left(Z_i = 0, X_i; T_j, M_j \right)
        \end{aligned}
    \end{equation}
    
    where $g^\text{post}_j := g_j | \mathcal{D}$ is the posterior assignment function for tree $j$. For ease of notation, let $\tau^{\text{BART}}_i:=\tau^{\text{BART}} \left( X_i\right)$. Let $I_\mathcal{O} \subseteq \{ 1,\dots, n \}$ be the indices of individuals in the stratified cluster sample that belong to the subpopulation group $\mathcal{O}$. Then the estimated distribution of the CATE for $\mathcal{O}$ is defined as $\frac{1}{|I_\mathcal{O}|} 
    \sum_{i \in I_\mathcal{O}} \tau_i^\text{BART}$. Likewise, the estimated distribution for the ATE is $\frac{1}{n} 
    \sum_{i \in [n]} \tau_i^\text{BART}$.
    
\end{enumerate}

Point estimates and standard errors come from the expected value and variance that's calculated based on the estimated distribution $\frac{1}{|I_\mathcal{O}|} \sum_{i \in I_\mathcal{O}} \tau_i^\text{BART}$. This is the standard application of BART and its variants for causal inference, as seen in \cite{hahn2020bayesian, hill2011bayesian}. \ 
    
    In the original NLSM study, a variant of BART (Bayesian causal forests \cite{hahn2020bayesian}) was used to calculate CATEs for subpopulations of varying School Achievement Level.

\item \textbf{BART with poststratification after imputing a point estimate of Prev-GPA for every poststratification cell (BARP-I) \cite{bisbee2019barp}}

\begin{enumerate}
    \item Fit the BART model in \ref{eqn:bart} to the stratified cluster sample $\mathcal{D}$.

    \item Let $c$ be a cell for poststratification matrix $M$ and $(me_c,g_c, re_c, s_c, sa_c, mc_c)$ be the corresponding covariate of School-Achievement $\times$ Minority-Composition $\times$ School $\times$ Gender $\times$ Race/Ethnicity $\times$ Maternal-Education for cell $c$. Define $\hat{v_c}$ to be a point estimate of Prev-GPA for poststratification cell $c$, based off the sample $\mathcal{D}$. There is no restriction on how to calculate $\hat{v_c}$ -- A straightforward approach would be to fit a hierarchical regression model with Prev-GPA as the outcome and the covariates being School-Achievement $\times$ Minority-Composition $\times$ School $\times$ Gender $\times$ Race/Ethnicity $\times$ Maternal-Education, and then take the posterior mean of the linear predictor evaluated at $(me_c,g_c, re_c, s_c, sa_c, mc_c)$ as the point estimate for $\hat{v_c}$. 
    
    \item Infer the posterior predictive distributions $$Y^{\text{rep},n}_{1,c} \sim p(y^*|z=1,v=\hat{v_c},me_c,g_c, re_c, s_c, sa_c, mc_c ,\mathcal{D})$$ $$Y^{\text{rep},n}_{0,c} \sim p(y^*|z=0,v=\hat{v_c},me_c,g_c, re_c, s_c, sa_c, mc_c ,\mathcal{D})$$ with the model in Equation $\ref{eqn:bart}$. Do this for all cells $c$ in poststratification matrix $M$. For posteriors of poststratification cells with no observation in $\mathcal{D}$, we use the out-of-sample capabilities of \texttt{dbarts} to infer the posterior predictive distributions $Y^{\text{rep},n}_{1,c}$ and $Y^{\text{rep},n}_{0,c}$.
    
    \item Use the poststratified distribution $$\frac{\sum_{c =1}^J N_c \left( Y^{\text{rep},n}_{1,c} - Y^{\text{rep},n}_{0,c} \right)}{\sum_{c' = 1}^J N_{c'}}$$ to get the estimated distribution for $\tau_\text{ATE}$, and $$\mathbb{E}\left(\frac{\sum_{c =1}^J N_c \left( Y^{\text{rep},n}_{1,c} - Y^{\text{rep},n}_{0,c} \right)}{\sum_{c' =1}^J N_{c'}} | \mathcal{D} \right)$$ as the point estimate for $\tau_\text{ATE}$. 
    
    In general, let $I_\mathcal{O}$ be the index set containing poststratification cells for the subpopulation $\mathcal{O}$. This subpopulation does not need to be observed in our population. Use Equation \ref{eqn:ate_estimator_}
    
        \begin{equation}
        \begin{aligned}
    \hat{\tau}_{\text{CATE},\mathcal{O}} &:=\frac{\sum_{j \in I_\mathcal{O}} N_c \left( Y^{\text{rep},n}_{1,c} - Y^{\text{rep},n}_{0,c} \right)}{\sum_{c' \in I_\mathcal{O}} N_{c'}}
        \end{aligned}
        \label{eqn:ate_estimator_}
    \end{equation} 
    
     to get the estimated distribution for $\tau_{\text{CATE},\mathcal{O}}$, and $\mathbb{E}\left( \hat{\tau}_{\text{CATE},\mathcal{O}} | \mathcal{D} \right)$ as the point estimate for $\tau_{\text{CATE},\mathcal{O}}$ and $\mathbb{V}\left( \hat{\tau}_{\text{CATE},\mathcal{O}} | \mathcal{D} \right)$ as the accompanying variance.

\end{enumerate}

\item \textbf{MRP after imputing a point estimate of Prev-GPA for every poststratification cell (MRP-I)} 

\begin{enumerate}

    \item Using $\mathcal{D}$, fit a hierarchical linear model $\mathcal{M}_\text{Post-GPA}$ with Post-GPA as the outcome, and Prev-GPA as a covariate. Assume the outcome is a truncated normal in $(0,4.33)$. The hierarchical modeling package \texttt{brms} \cite{burkner2018brms,burkner2017brms} will be used for fitting $\mathcal{M}_\text{Post-GPA}$. For more details on the specification of $\mathcal{M}_\text{Post-GPA}$, refer to Section 5 of this manuscript.
    
    \item Let $c$ be a cell for poststratification matrix $M$ and $(me_c,g_c, re_c, s_c, sa_c, mc_c)$ be the corresponding covariate of School-Achievement $\times$ Minority-Composition $\times$ School $\times$ Gender $\times$ Race/Ethnicity $\times$ Maternal-Education for cell $c$. For every poststratification cell $c$, define $\hat{v_c}$ to be a point estimate of Prev-GPA for poststratification cell $c$, based off the sample $\mathcal{D}$
    
    \item Using $\mathcal{M}_\text{Post-GPA}$, infer the posterior predictive distributions
    $$Y^{\text{rep},n}_{1,c} \sim p(y^*|z=1,v=\hat{v_c},me_c,g_c, re_c, s_c, sa_c, mc_c \mathcal{D})$$ $$Y^{\text{rep},n}_{0,c} \sim p(y^*|z=0,v=\hat{v_c},me_c,g_c, re_c, s_c, sa_c, mc_c ,\mathcal{D})$$ for every poststratification cell $c$ in $M$. For posteriors of poststratification cells with no observation in $\mathcal{D}$, we use \texttt{brms} to infer the posterior predictive distributions $Y^{\text{rep},n}_{1,c}$ and $Y^{\text{rep},n}_{0,c}$ (More details on this is in Subsection 5.2.3).
    
    \item Use the poststratified distribution $$\frac{\sum_{c =1}^J N_c \left( Y^{\text{rep},n}_{1,c} - Y^{\text{rep},n}_{0,c} \right)}{\sum_{c' = 1}^J N_{c'}}$$ to get the estimated distribution for $\tau_\text{ATE}$, and $$\mathbb{E}\left(\frac{\sum_{c =1}^J N_c \left( Y^{\text{rep},n}_{1,c} - Y^{\text{rep},n}_{0,c} \right)}{\sum_{c' =1}^J N_{c'}} | \mathcal{D} \right)$$ as the point estimate for $\tau_\text{ATE}$. 
    
   In general, let $I_\mathcal{O}$ be the index set containing poststratification cells for the subpopulation $\mathcal{O}$. This subpopulation does not need to be observed in our population. Use Equation \ref{eqn:ate_estimator_} to get the estimated distribution for $\tau_{\text{CATE},\mathcal{O}}$, and $\mathbb{E}\left( \hat{\tau}_{\text{CATE},\mathcal{O}} | \mathcal{D} \right)$ as the point estimate for $\tau_{\text{CATE},\mathcal{O}}$ and $\mathbb{V}\left( \hat{\tau}_{\text{CATE},\mathcal{O}} | \mathcal{D} \right)$ as the accompanying variance.
     
\end{enumerate}

In our simulation studies, Step (b) for MRP-I and BARP-I is performed by first using the observed sample $\mathcal{D}$ to fit a hierarchical regression model $\mathcal{M}_\text{Prev-GPA}$ with Prev-GPA as the outcome and the covariates being School-Achievement $\times$ Minority-Composition $\times$ School $\times$ Gender $\times$ Race/Ethnicity $\times$ Maternal-Education. $\mathcal{M}_\text{Prev-GPA}$ in BARP-I is a BART specification and $\mathcal{M}_\text{Prev-GPA}$ in MRP-I is a hierarchical regression model specification. For a more detailed specification of $\mathcal{M}_\text{Prev-GPA}$ for MRP-I in our simulation studies, refer to Section 5 of this manuscript. Once $\mathcal{M}_\text{Prev-GPA}$ is fit, the posterior predictive distribution of Prev-GPA for every poststratification cell $c$ based off $\mathcal{M}_\text{Prev-GPA}$ has its mean set as $\hat{v}_c$. %

Posterior predictive distributions for the outcome Post-GPA are used for methods BARP-I and MRP-I, since they are inferring $Y^{\text{rep},n}_{1,c}, Y^{\text{rep},n}_{0,c}$ for poststratification cells $c$ that are not observed in $\mathcal{D}$. Both these methods are able to calculate CATEs for subpopulations not observed in $\mathcal{D}$. On the other hand, OLS, SVY, BART are only able to estimate CATEs observed in $\mathcal{D}$. 

As motivated in the previous subsection, including poststratification-style estimators such as BARP-I and MRP-I is sensible as we're provided a poststratification matrix $M$. The current challenge of performing MRP in our simulation setting is that, if one wants to include non-census variable Prev-GPA as a covariate in the outcome model for MRP, then atleast one point estimate of Prev-GPA will have to be used for every poststratification cell in $M$. BARP-I and MRP-I performs this by imputing an estimated Prev-GPA quantity for every poststratification cell. This is because $M$ is a poststratification matrix of all combinations for the discrete covariates School-Achievement $\times$ Minority-Composition $\times$ School $\times$ Gender $\times$ Race/Ethnicity $\times$ Maternal-Education, but there exists no discretization of continuous covariate Prev-GPA where there is a poststratification matrix of Prev-GPA $\times$ School-Achievement $\times$ Minority-Composition $\times$ School $\times$ Gender $\times$ Race/Ethnicity $\times$ Maternal-Education. 

\end{enumerate}

This challenge of having truncated continuous covariate Prev-GPA with no poststratification matrix for a discretization of Prev-GPA motivates the application of the variant of MRP, MRP-MI. From a high level, MRP-MI estimates poststratification frequencies and  uses these estimated frequencies to perform poststratification. This accounts for the full distribution of Prev-GPA in every poststratification cell $c$, unlike what MRP-I and BARP-I does by taking Prev-GPA point estimate $\hat{v}_c$.

\section{Multilevel Regression and Poststratification with Multiple Imputation}

We use the MRP variant, Multilevel Regression and Poststratification with Multiple Imputation (MRP-MI), as a viable framework for estimating treatment effects for our simulation study.  It can be applied when the modeler is given a poststratification matrix $\tilde{M} \in \mathbb{R}^{J \times K}$ for the $K$-dimensional categorical covariate vector $\tilde{C} \in \mathbb{R}^K$ and wants to also poststratify by a continuous covariate $\tilde{V} \in \mathbb{R}$ but that continuous covariate does not have population counts in poststratification matrix $\tilde{M}$.

As mentioned in subsection 1.4, using Post-GPA - Prev-GPA as the outcome is not possible, thus we have to include Prev-GPA as a covariate and Post-GPA as the outcome when modeling and estimating treatment effects in our causal inference problem (Figure \ref{fig:dag}). In our simulation study, the poststratification matrix $M$ does not contain population counts for strata of Prev-GPA, but we'd still like to poststratify by Prev-GPA $\times$ School-Achievement $\times$ Minority-Composition $\times$ School $\times$ Gender $\times$ Race/Ethnicity $\times$ Maternal-Education and MRP-MI is able to do this. The three steps below outline the framework for MRP-MI. More detail and motivation for model specifications relevant to our simulation study are in Section 5. \texttt{brms} is used to fit the Post-GPA and Prev-GPA models used in the MRP-MI framework.

\begin{enumerate}
    \item \textbf{Model fitting step:} Fit a hierarchical regression model $\mathcal{M}_\text{Post-GPA}$ to the observed sample $\mathcal{D} = (V_i, C_i, Z_i, Y_i)_{i=1}^n$, where $V_i$ is Prev-GPA, $C_i = \left(\text{ME}_i,\text{G}_i,\text{RE}_i, \text{School}_i,\text{SA}_i,\text{MC}_i \right)$ is the categorical covariate, and $Z_i$ is the treatment variable and $Y_i$ is the Post-GPA outcome for individual $i$. For more details on the specification of $\mathcal{M}_\text{Post-GPA}$ in our simulation study, refer to Equation \ref{eqn:post_gpa_model} in Section 5 of this manuscript. 
    
    Additionally, fit a hierarchical regression model $\mathcal{M}_\text{Prev-GPA}$ to the observed sample $\mathcal{D}$, where Prev-GPA $V_i$ is the outcome and the covariates are $C_i$. For more details on the specification of $\mathcal{M}_\text{Prev-GPA}$, refer to Equation \ref{eqn:prev_gpa_model} in Section 5 of this manuscript.

    \item \textbf{Sampling for Post-GPA PPD step:} Let $p(y^* | c, z, v^*, \mathcal{D})$ be the posterior predictive distribution of Post-GPA for the new datapoint $(c,z,v^*)$ based on $\mathcal{M}_\text{Post-GPA}$. Let $p(v^* | c, \mathcal{D})$ be the posterior predictive distribution of Prev-GPA for the new datapoint $(c,z)$ based on $\mathcal{M}_\text{Prev-GPA}$. Define $Y^{\text{rep},n}_{z,c} \sim p(y^* | c, z, \mathcal{D})$, where 
    
    \begin{equation}
        \begin{aligned}
            p(y^* | c, z, \mathcal{D}) &= \int p(y^*| c, z , v^* , \mathcal{D}) p(v^* | c, z, \mathcal{D}) dv^* \\
            & = \int p(y^*| c, z , v^*,\mathcal{D} ) p(v^* | c, \mathcal{D}) dv^*, \text{ since $\mathcal{M}_\text{Prev-GPA}$ does not have $z$ as a covariate}
        \end{aligned}
        \label{eqn:integratingitout}
    \end{equation}

    Hence $p(y^* | c, z, \mathcal{D})$ is the posterior predictive distribution of Post-GPA in poststratification cell $c$, after integrating out the posterior predictive distribution of Prev-GPA for cell $c$. To retrieve samples from $p(y^* | c, z, \mathcal{D})$, the below two-step procedure performs it:
    
\begin{minipage}{0.94 \textwidth}
    \begin{algorithm}[H]
    \SetAlgoLined
    \KwResult{Return 1000 samples $(y^{*j}_{z,c})_{j=1}^{1000}$ of $Y^{\text{rep},n}_{z,c} \sim p(y^* | c, z, \mathcal{D})$ for every combination of $(z,c)$  }
    
     $m = 1000$\;
    \For{ all combinations of $(c,z)$}{
         \For{$j \in \{1, \dots, m\}$ }{
            Sample $v^{*j}_c$ from $p(v^* | c, \mathcal{D})$\;
            Sample $y^{*j}_{z,c}$ from $p(y^* | c,z,v^{*j}_c, \mathcal{D})$
         }
     }

 \caption{MRP-MI}
\end{algorithm}
\end{minipage}

For poststratification cells $c$ not observed in $\mathcal{D}$, the out-of-sample capabilities of \texttt{brms} are used to sample $v^{*j}_c$ and $y^{*j}_{z,c}$. More details on how this is done can be found in Section 5.2.3.

The above algorithm uses the posterior predictive distribution of $\mathcal{M}_\text{Prev-GPA}$ to impute Prev-GPA 1000 times in every poststratification cell $c$, hence the acronym MI (multiple imputation). Note that the above algorithm actually generates the samples $\left(\theta^{[j]}_v, v^{*j}_c, \theta_y^{[j]}, y^{*j}_{c,z} \right)_{j=1}^{1000}$, where $\left(\theta^{[j]}_v \right)_{j=1}^{1000}$ and $\left(\theta^{[j]}_y \right)_{j=1}^{1000}$ are posterior samples of the Prev-GPA and Post-GPA model respectively. However, this way of sampling introduces a dependency structure into the joint random variable (after conditioning on sample $\mathcal{D}$) $$\left(Y^{\text{rep},n}_{0,1},\dots,Y^{\text{rep},n}_{0,J},Y^{\text{rep},n}_{1,1},\dots,Y^{\text{rep},n}_{1,J} \right)$$ To make samples from this joint random variable conditionally mutually independent (when conditioned on the dataset $\mathcal{D}$), we have to use mutually exclusive posterior draws of $\left( \theta_v^{[j]}, \theta_y^{[j]} \right)_{j=1}^{1000}$. This means using the first $\frac{1000}{2J}$ draws for $Y^{\text{rep},n}_{0,1}$,  the second $\frac{1000}{2J}$ draws for $Y^{\text{rep},n}_{0,2}$ and so on. This wouldn't be feasible if $J$ is large, which in MRP applications it usually is.

     \item \textbf{Poststratification step:} The estimated distribution for CATE of subpopulation $\mathcal{O}$ based on dataset $\mathcal{D}$ is 

\begin{equation}
    \begin{aligned}
\hat{\tau}_{\text{CATE},\mathcal{O}} &=\frac{\sum_{c \in I_ \mathcal{O}} N_c \left( Y^{\text{rep},n}_{1,c} - Y^{\text{rep},n}_{0,c} \right)}{\sum_{c' \in I_ \mathcal{O}} N_{c'}}
    \end{aligned}
    \label{eqn:ate_estimator_ourmethod}
\end{equation}

Based off the above algorithm, we get 1000 samples of the estimator defined in Equation \ref{eqn:ate_estimator_ourmethod} through 

\begin{equation}
    \begin{aligned}
        \left(\frac{\sum_{c \in I_ \mathcal{O}} N_c \left( y_{1,c}^{*j} - y_{0,c}^{*j} \right) }{\sum_{c' \in I_ \mathcal{O}} N_{c'} }  \right)_{j=1}^{1000}
    \end{aligned}
    \label{eqn:ate_estimator_ourmethod_samples}
\end{equation}

\end{enumerate}

We have the below proposition for the framework MRP-MI when it's used to calculate the point estimate of the CATE for subpopulation $\mathcal{O}$. The proof of this proposition is in Supplementary Material A.

\begin{proposition}{Let $g_z(c,v^*) = \int y^* p(y^* |z,c,v^*,\mathcal{D}) dy^*$ be the expected value of the posterior predictive distribution of Post-GPA model $\mathcal{M}_\text{Post-GPA}$ under treatment $Z=z$, categorical covariate $C = c$ and Prev-GPA $V = v^*$. Calculating $\mathbb{E}\left(\hat{\tau}_{\text{CATE}, \mathcal{O}} | \mathcal{D} \right)$ is integrating $g_1(c,v^*) - g_0(c,v^*)$ with the continuous probability measure defined by $\left\{ \frac{N_c}{\sum_{c' \in I_\mathcal{O} } N_{c'}} p(v^* |c,\mathcal{D}) dv^* \right\}_{(c,v^*) \in I_\mathcal{O} \times (0,4.33)}$.}

\begin{proof}
See Supplementary Material A.
\end{proof}

\label{prop:poststratprobmeasure}
\end{proposition}

If we consider $g_z(c,v^*)$ as an estimate for $\mathbb{E} \left( Y (z) | C = c, V = v^* \right)$ then Proposition 1 shows that $\mathbb{E}\left(\hat{\tau}_{\text{CATE},\mathcal{O}} |\mathcal{D} \right)$ is poststratifying $g_1(c,v^*) - g_0(c,v^*)$ with the poststratification weights defined by
\begin{equation}
    \begin{aligned}
        \mathbb{P} \left( C = c,V = v^* | (C,V) \in \mathcal{O} \right) &= \mathbb{P} \left( V = v^* |C = c, (C,V) \in \mathcal{O} \right) \mathbb{P} \left( C = c |(C,V) \in \mathcal{O} \right)\\ 
        &= p(v^* |c,\mathcal{D}) \frac{N_c}{\sum_{c' \in I_\mathcal{O} } N_{c'}}
    \end{aligned}
\end{equation}

This is because $\left\{ p(v^* |c,\mathcal{D}) \frac{N_c}{\sum_{c' \in I_\mathcal{O} } N_{c'}} \right\}_{(c,v^*) \in I_\mathcal{O} \times (0,4.33)}$ can be viewed as the poststratification weights for all combinations of Prev-GPA $\times$ School-Achievement $\times$ Minority-Composition $\times$ School $\times$ Gender $\times$ Race/Ethnicity $\times$ Maternal-Education when Prev-GPA is discretized very fine. 

In contrast to the methods BARP-I and MRP-I which impute Prev-GPA just once for every poststratification cell, we see that MRP-MI (with $\mathcal{M}_\text{Post-GPA}$ and $\mathcal{M}_\text{Prev-GPA}$ specified in Section 5) accounts for the full distribution of Prev-GPA in every poststratification cell, $p(v^* | c, \mathcal{D})$, by integrating it out as seen in Equation \ref{eqn:integratingitout}. The trade-off of accounting for the whole distribution of Prev-GPA in every poststratification cell is that the computational cost of getting 1000 samples for $\hat{\tau}_{\text{CATE},\mathcal{O}}$ is significantly higher.

\section{Simulation results}

A detailed outline of the data generating process for our simulation study that modeled the NSLM study is provided in this section. We first generate a finite population $P$ from a superpopulation, then utilize a stratified sampling scheme to generate the observed sample $\mathcal{D}$.

Simulation results of calculating CATEs for subpopulations of varying sizes is presented and the six treatment effect estimation methods are compared. We evaluate the efficacy of each estimation method with various metrics that measure the quality of point estimation, standard error and uncertainty interval coverage.

The individuals in NSLM were recruited through a stratified cluster sample, where the clusters are schools. Stratification was performed on school-level categorical covariates, and the outcome variable of interest was Post-GPA. In the end, around 70 high-schools had agreed to participate in the study, and randomization of the growth mindset intervention was performed at the individual level of the stratified cluster sample. Despite the population of US public high-schools being $12,000$, NSLM returned a highly representative sample. The estimated intervention effectiveness on GPA from the NSLM stratified cluster sample in turn was generalizable to the target population of $9^{\text{th}}$ grade public high-school students in the United States.

\texttt{R} \cite{R2020} was the computing language used for the simulation studies and the \texttt{ggplot2} package \cite{wickhamggplot} was used to generate the plots in this manuscript. \texttt{brms} was the library used for hierarchical model fitting. \cite{burkner2018brms,burkner2017brms,stan2017}.
 
\subsection{Data generating process in simulation setup for potential outcomes in finite population $P$}

Let $i$ be the index of an individual, $k$ be index of the school, $s$ be index of the stratum. $V_{iks}$ is the previous-GPA for student $i$ in school $k$ in stratum $s$. $Y_{iks}$ is the post-GPA and $Z_{iks}$ is the treatment-control indicator.

Page 31 of the appendix of the NSLM study \cite{yeager2019national} contained descriptive statistics of the student-level and school-level covariates seen in Figure \ref{fig:dag}. These descriptive statistics were used to define the coefficients used in the simulated version of the NSLM study as seen below. The DAG in Figure \ref{fig:dag} represents the data generating process. 

The sizes of each stratum, $M_s$, are $(2806,3040,2570,2239,566)$ and the corresponding minority composition and school achievement levels are:

\begin{table}[ht]
\centering
\begin{tabular}{rrrrr}
  \hline
 Strata & Minority Composition & School Achievement Level & Number of schools $M_s$ \\ 
  \hline
1 & Both & Low & 2806 \\ 
  2 & Low & Medium & 3040 \\ 
  3 & High & Medium & 2570 \\ 
  4 & Low & High & 2239 \\ 
  5 & High & High & 566 \\ 
   \hline
\end{tabular}

\caption{Number of schools within each of the 5 strata defined by Minority Composition level and School Achievement level.}
\label{tab:school_strata}
\end{table}

Generate the unobserved school-level noises $U^\text{School}_{ks}, T^\text{School}_{ks}$ just once:

\begin{equation}
    \begin{aligned}
        U^\text{School}_{ks} &\overset{\text{i.i.d}}{\sim} \mathcal{N} (0, 0.04^2)\\
        T^\text{School}_{ks} &\overset{\text{i.i.d}}{\sim} \mathcal{N} (0, 0.04^2)
    \end{aligned}
\end{equation}

$U^\text{School}_{ks}, T^\text{School}_{ks}$ are on the same scale of the treatment effects seen in Extended Data Figure 2 in the original NSLM study \cite{yeager2019national}.

\subsubsection{Data generating process for $P$}

For each stratum, $s$, let $M_{ks} \sim \text{Poisson}(200)$ be the school size for school $k$ within stratum $s$. Then for $i \in [M_{ks}]$, we generate the covariates along with previous and post GPA conditionally independently:

\begin{equation}
    \begin{aligned}
        \text{G}_{iks} &\sim \text{Bernoulli}(0.49)\\
        V_{iks} &\sim \mathcal{N}_{(0,4.33)}\left(\mu_X[s],\sigma_X[s]\right) \\
        \text{RE}_{iks} &\sim \text{Cat}\left( p_{\text{RE}}[s] \right)\\
        \text{ME}_{iks} &\sim \text{Bernoulli}\left( p_{\text{ME}}[s] \right)\\
        Y_{iks} (Z_{iks}) |\text{G}_{iks}, V_{iks}, \text{RE}_{iks}, \text{ME}_{iks},U^\text{School}_{ks},T^\text{School}_{ks}  &\sim \mathcal{N}_{(0,4.33)}(\text{Max}(0,\text{Min}(4.33,V_{iks} + \tau_{\text{SA}}[s]Z_{iks}\\
        &+ \tau_{\text{MC}}[s]Z_{iks} + \tau_{\text{G}}[\text{G}_{iks}]Z_{iks} + \tau_{\text{RE}}[\text{RE}_{iks}]Z_{iks}\\
        &+ \tau_{\text{ME}}[\text{ME}_{iks}]Z_{iks} + U^\text{School}_{ks}Z_{iks} + T^\text{School}_{ks})),0.6  )
    \end{aligned}
    \label{eqn:dgprocess}
\end{equation}

$\text{G}_{iks}$ is the gender indicator, $\text{RE}_{iks}$ is the race/ethnicity, $\text{ME}_{iks}$ is the maternal education indicator, $V_{iks}$ is the previous GPA, $Y_{iks}$ is the post-GPA and $Z_{iks}$ is the treatment indicator. The coefficient vectors $\mu_X, \sigma_X, p_\text{RE}, p_\text{ME}, \tau_{\text{SA}}, \tau_{\text{MC}}, \tau_{\text{G}}, \tau_{\text{RE}}, \tau_{\text{ME}}$ are defined in the Supplementary Material C.

We can assume that this data generating process for $P = \left(Y_m(0),Y_m(1),X_m \right)_{m=1}^{N_P}$ which is around 2.2 million individuals is from the data generating process of $i.i.d$ draws from a superpopulation $\left(Y(0), Y(1), X \right)$. This is because fixing the strata sizes $M_s$ is equivalent to sampling from a 5-dimensional categorial variable with mean being equal to 5-dimensional vector $\left( \frac{M_s}{\sum_{s'=1}^5 M_{s'}} \right)_{s=1}^5$ in the asymptotic regime.

\subsubsection{Sampling from the finite population $P$}

The population $P$ is generated once and then a high-quality stratified cluster sample $\mathcal{D}$ of around $12,000$ individuals is taken. The stratified sampling design we implement closely resembles the experimental design used in the NSLM data \cite{yeager2019national}. This is done in the following steps:

\begin{enumerate}
    \item \textbf{Stratified sample of clusters (schools):} Sample $(28,34,32,19,27)$ schools randomly (without replacement) from each of the 5 stratum respectively. This returns students in 140 schools.
    
    \item \textbf{Subsample clusters with response probabilities on units in clusters:} Select $\text{Poissson}\left( \frac{1}{65} \right)$ schools from the 140. Every student in the population has response probability $\text{invlogit}^{-1}(V_{iks})$, the inverse-logit of the previous GPA for that student. Define this size $n$ sample as $\mathcal{D}$.
\end{enumerate}

The individual response probability $\text{invlogit}^{-1}(V_{iks})$ is $92$ percent. This agrees with the high response rate in the NSLM study \cite{yeager2019national}.

Generating $P$ is done using the \texttt{R} package \texttt{DeclareDesign} \cite{blair2019declaredesign}. \texttt{DeclareDesign} assigns $Z_i$ at the the finite population level. That is, the assignment mechanism at the finite population level is complete randomization of $\left( Z_i \right)_{i=1}^{N_P}$, giving $\sum_{i=1}^{N_P} Z_i = \frac{N_P}{2}$. Because $\mathcal{D}$ is a very small sample of the finite population $P$ ($\mathcal{D}$ is 0.5 percent of $P$), the $Z_i$ in $\mathcal{D}$ are approximately independent Bernoulli$(0.5)$ samples. Hence the assignment mechanism for $\mathcal{D}$ can be thought of as a 
Bernoulli randomization---for every individual in $\mathcal{D}$, we independently assignment them a 50 percent probability of being in the treatment group. As expected, the stratified cluster sample $\mathcal{D}$ still maintains close to a $50-50$ randomization between treatment and control.

It's clear to see that identifiability assumptions 1 - 4 are satisfied for $\mathcal{D}$. SUTVA and Consistency are satisfied through the data generating process for $P$, Ignorability is satisfied through the assignment mechanism for $\left(Z_i \right)_{i=1}^{N_P}$, and through the assignment mechanism we have $\mathbb{P}\left(Z_i = 1 | X_i \right) = 0.5$ thus satisfying the Overlap condition. It's important to note that, even though we generate a finite population $P$ and then sample from it, we calculate CATEs and ATE at the superpopulation level.%

Generated $\mathcal{D}$ was fairly representative of the target population $P$. The Tipton Generalization Index \cite{tipton2014generalizable} for $\mathcal{D}$ along with the subgroups we analyze in each simulation iteration was calculated to be near $0.9$ (calculated using the \texttt{generalize} package \cite{ackerman2019implementing} with random forests being the selection method). This matches the high Generalization Index of the sample in the original NSLM study.

\subsubsection{Superpopulation estimands: G-formula for superpopulation ATE and CATEs based on the data generating process}

As the identifiability assumptions are satisfied for the data generating process of $P$, we can derive formulas for $\tau_{\text{CATE},\mathcal{O}}$ for any subpopulation $\mathcal{O}$.

Given that $U_{ks}^\text{School}=u_{ks},T_{ks}^\text{School}=t_{ks}$ for every stratum $s$ and school $k$, let $Y(z)$ be a random sample from the population defined in Equation \ref{eqn:dgprocess}, under treatment $Z=z$. The G-formula \cite{hernanchapman} applied to the data-generating process recovers the true ATE:

\begin{equation}
\begin{aligned}
    \mathbb{E}\left(Y(z) \right) =& \sum_{s} 
    \sum_{k} \mathbb{E}(Y(z) | z, s, u_{ks},t_{ks}) \frac{M_s}{\sum_{s'}M_{s'}} \frac{1}{M_s}\\
     =& \sum_{s} \sum_{k} \sum_{me} \sum_{g} \sum_{re} \left(\int_0^{4.33} \mathbb{E}(Y(z) | z, s, me, g, re, v, u_{ks},t_{ks}) p(v|s) dv \right)\\
    & \cdot p(re|s) p(me|s) p(g) \frac{M_s}{\sum_{s'}M_{s'}} \frac{1}{M_s}
\end{aligned}
\end{equation}

Applying the G-formula for the whole target population gives $\tau_{\text{ATE}} := \mathbb{E}\left(Y(1)\right) - \mathbb{E}\left(Y(0)\right) = 0.126$ (rounded to 3 decimal places).

Note that $ \mathbb{E}\left(Y(z) | z, s, me, g, re, k\right)=\int_0^{4.33} \mathbb{E}\left(Y(z) | z, s, me, g, re, v, u_{ks},t_{ks}\right) p_{{V}|\text{S}}(v|s) dv$ is the expected value of Post-GPA under $Z=z$ for poststratification cell School-Achievement $\times$ Minority-Composition $\times$ Gender $\times$ Race/Ethnicity $\times$ Maternal-Education $\times$ School $= (s,g,re,me,k)$, after integrating out Prev-GPA. Hence summing $\mathbb{E}\left(Y(z) | z, s, me, g, re, k\right) \mathbb{P} \left( s, me, g, re, k | X \in \mathcal{O} \right)$ over all the poststratification cells in subpopulation $\mathcal{O}$ returns $\mathbb{E} \left( Y(z) | X \in \mathcal{O} \right)$, which in turn will result in the CATE $\tau_{\text{CATE}, \mathcal{O}}$.

\subsection{Hierachical model specifications for $\mathcal{M}_\text{Prev-GPA}$ and $\mathcal{M}_\text{Post-GPA}$ which are used in BARP-I, MRP-I, MRP-MI}

\begin{figure}[ht]
    \centering
    \includegraphics[width=1\textwidth]{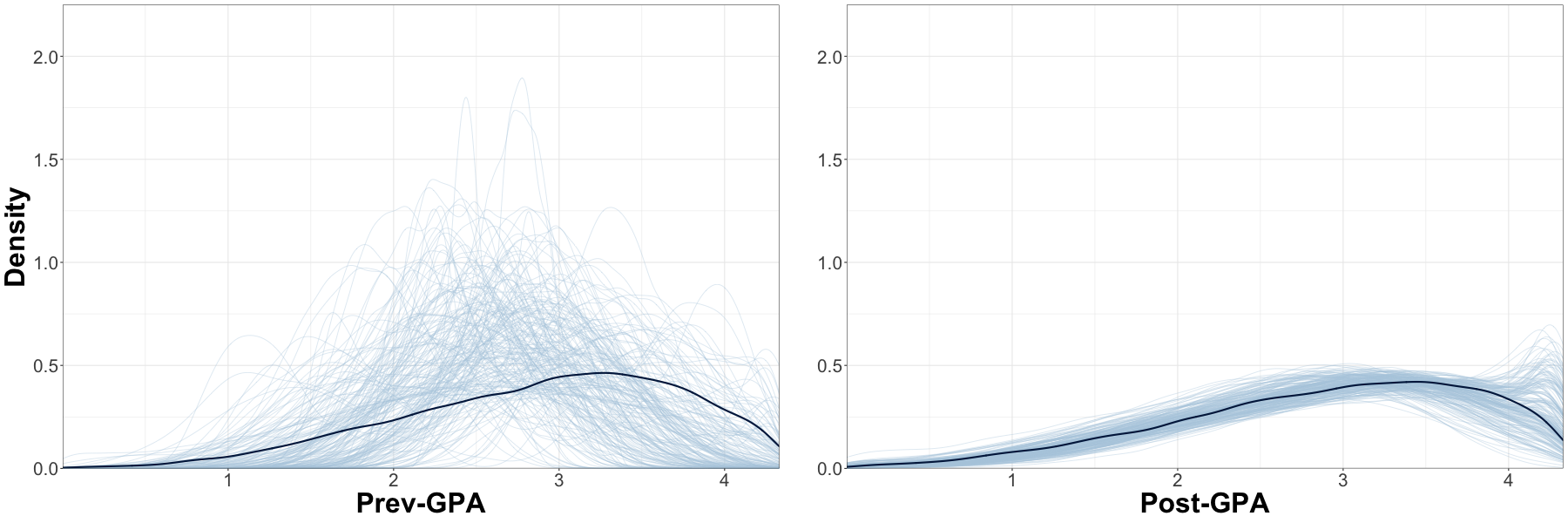}
    \caption{Prior predictive checks for $\mathcal{M}_\text{Prev-GPA}$ and $\mathcal{M}_\text{Post-GPA}$, models which are used in MRP-MI, MRP-I and BARP-I. (Left) The dark-blue line corresponds to the density of Prev-GPA observations in $\mathcal{D}$. The 200 light-blue lines correspond to the densities of 200 prior predictive draws of Prev-GPA for individuals in $\mathcal{D}$. (Right) The dark-blue line corresponds to the density of Post-GPA observations in $\mathcal{D}$. The 200 light-blue lines correspond to the densities of 200 prior predictive draws of Post-GPA for individuals in $\mathcal{D}$.
    }
    \label{fig:ppc}
\end{figure}

Prior predictive checks \cite{bayesplot2021, gabry2019visualization, gelman2013bayesian} as seen in Figure \ref{fig:ppc} were used to calibrate the prior specifications of the hierarchical models $\mathcal{M}_\text{Prev-GPA}$ and $\mathcal{M}_\text{Post-GPA}$. Fitted with the package \texttt{brms}, both hierachical models assume a truncated normal outcome in the range $(0,4.33)$, which matches the range that Prev-GPA and Post-GPA are in.

\subsubsection{Previous-GPA model $\mathcal{M}_\text{Prev-GPA}$}

The Prev-GPA model is fit to the stratified cluster sample $\mathcal{D}$ (a new fit happens for every simulation iteration since a new $\mathcal{D}$ is sampled from $P$). 

For the $i^{th}$ individual, define $\alpha_{j[i]}^\text{RE}$, $\alpha_{j[i]}^\text{School}$, $\alpha_{j[i]}^\text{MC}$, $\alpha_{j[i]}^\text{SA}$, $\alpha_{j[i]}^\text{SA,MC}$ to be their unique random intercepts for the covariates Race/Ethnicity, School, Minority Composition, School Achievement, Minority Composition $\times$ School Achievement respectively.

The outcome is modeled as:

\begin{equation}
\begin{aligned}
V_i|\text{ME}_i,\text{G}_i,\alpha^\text{RE}_{j}, \alpha^\text{School}_{j},\alpha^\text{MC}_{j},\alpha^\text{SA}_{j},\alpha^\text{SA,MC}_{j} &\sim \mathcal{N}_{(0,4.33)}\left(\beta_0 + \beta_\text{ME}\text{ME}_i + \beta_\text{G}\text{G}_i + \right. \\
&\left. \alpha^\text{RE}_{j[i]} + \alpha^\text{School}_{j[i]} + \alpha^\text{MC}_{j[i]} + \alpha^\text{SA}_{j[i]} + \alpha^\text{SA,MC}_{j[i]}, \sigma^2 \right)
\end{aligned}
\label{eqn:prev_gpa_model}
\end{equation}

Covariates $=$ RE, School, MC, SA, and  $\text{SA,MC}$ have atleast 3 levels and have their priors modeled as:

\begin{equation}
\begin{aligned}
\alpha_{j}^\text{Covariates} | \sigma^\text{Covariates} &\overset{\text{i.i.d}}{\sim} \mathcal{N} \left(0,(\sigma^\text{Covariates})^2 \right)\\
\sigma^\text{Covariates} &\sim \mathcal{N}_+ \left(0,0.25^2 \right)\\
\end{aligned}
\label{eqn:prev_gpa_model_randomeffect}
\end{equation}

Covariates $=$ ME and G are binary and have their priors (along with the global intercept) modeled as:

\begin{equation}
\begin{aligned}
\beta_\text{Covariates} &\sim \mathcal{N}\left(0, 0.25^2\right)\\
\beta_0 &\sim \mathcal{N}\left(2.7, 0.25^2\right)
\end{aligned}
\label{eqn:prev_gpa_model_betas}
\end{equation}

\subsubsection{Post-GPA model $\mathcal{M}_\text{Post-GPA}$}

The Post-GPA model is fit to the stratified cluster sample $\mathcal{D}$ (a new fit happens for every simulation iteration since a new $\mathcal{D}$ is sampled from $P$). 

For the $i^{th}$ individual, define $\gamma_{j[i]}^\text{RE}$, $\gamma_{j[i]}^\text{School}$, $\gamma_{j[i]}^\text{MC}$, $\gamma_{j[i]}^\text{SA}$, $\gamma_{j[i]}^\text{SA,MC}$ to be their unique random intercepts when $Z_i=1$, for the covariates Race/Ethnicity, School, Minority Composition, School Achievement, Minority Composition $\times$ School Achievement respectively.

The outcome is modeled as:

\begin{equation}
\begin{aligned}
Y_i|V_i,Z_i,\text{ME}_i,\text{G}_i,\alpha^\text{RE}_{j}, \alpha^\text{School}_{j},\alpha^\text{MC}_{j},\alpha^\text{SA}_{j},\alpha^\text{SA,MC}_{j} &\sim \mathcal{N}_{(0,4.33)}\left(\beta_0 + \text{Offset}(V_i) + \beta_\text{ME}\text{ME}_i + \beta_\text{G}\text{G}_i + \right. \\
&\left. \gamma_\text{ME}\text{ME}_i Z_i + \gamma_\text{G}\text{G}_i Z_i + \right. \\
&\left. \alpha^\text{RE}_{j[i]} + \alpha^\text{School}_{j[i]} + \alpha^\text{MC}_{j[i]} + \alpha^\text{SA}_{j[i]} + \alpha^\text{SA,MC}_{j[i]} + \right.\\
&\left. \gamma^\text{RE}_{j[i]}Z_i + \gamma^\text{School}_{j[i]}Z_i + \gamma^\text{MC}_{j[i]}Z_i + \gamma^\text{SA}_{j[i]}Z_i + \gamma^\text{SA,MC}_{j[i]}Z_i, \sigma^2 \right)
\end{aligned}
\label{eqn:post_gpa_model}
\end{equation}

Covariates $=$  RE, School, MC, SA, and  $\text{SA,MC}$ have atleast 3 levels and have their priors modeled as:

\begin{equation}
\begin{aligned}
\alpha_{j}^\text{Covariates} | \sigma^\text{Covariates} &\overset{\text{i.i.d}}{\sim} \mathcal{N} \left(0,(\sigma^\text{Covariates})^2 \right)\\
\sigma^\text{Covariates} &\sim \mathcal{N}_+ \left(0,0.125^2 \right)\\
\gamma_j^\text{Covariates} | \sigma_{\gamma}^\text{Covariates} &\overset{\text{i.i.d}}{\sim} \mathcal{N}(0, (\sigma_{\gamma}^\text{Covariates})^2)\\
\sigma_{\gamma}^\text{Covariates} &\sim \mathcal{N}_+(0, 0.125^2)\\
\text{Cor}(\gamma_j^\text{Covariates}, \alpha_{j}^\text{Covariates})  &\sim \text{Default prior in \texttt{brms}}
\end{aligned}
\label{eqn:post_gpa_model_Z}
\end{equation}

Covariates $=$ ME and G are binary and have their priors (along with the global intercept) modeled as:

\begin{equation}
\begin{aligned}
\beta_\text{Covariates} &\sim \mathcal{N}\left(0, 0.125^2\right)\\
\gamma_\text{Covariates} &\sim \mathcal{N}\left(0, 0.125^2\right)\\
\beta_0 &\sim \mathcal{N}\left(0, 0.125^2\right)
\end{aligned}
\label{eqn:post_gpa_model_betas}
\end{equation}

\subsubsection{Out-of-sample prediction with $\mathcal{M}_\text{Prev-GPA}$ and $\mathcal{M}_\text{Post-GPA}$}

The treatment effect estimation methods MRP-I and MRP-MI use $\mathcal{M}_\text{Prev-GPA}$ and $\mathcal{M}_\text{Post-GPA}$ to sample from the Post-GPA and Prev-GPA posterior predictive distributions $(Y_{z,c}^\text{rep,n}, V_{z,c}^\text{rep,n})$ of poststratification cells $c$. BARP-I uses $\mathcal{M}_\text{Prev-GPA}$ to sample from the Prev-GPA posterior predictive distribution $V_{z,c}^\text{rep,n}$ of poststratification cells $c$.

In our simulation studies, inferring posterior predictive samples of $(Y_{z,c}^\text{rep,n}, V_{z,c}^\text{rep,n})$ for a cell $c$ not observed in $\mathcal{D}$ requires sampling from new levels $j$ of the posteriors $\alpha_{j}^\text{School}$ and $\gamma_{j}^\text{School}$. This is because $\mathcal{D}$ is a stratified cluster sample consisting of around 70 schools (school is a cluster), leaving 11000 other unobserved schools in the target population. Sampling from the posterior of $\alpha_{j}^\text{School}$ for an unobserved school $j$ is done by sampling from a normal distribution with mean implied by the posteriors $\alpha_{j^*}^\text{School}$ of all the observed schools $j^*$, and standard deviation implied by the posterior distribution of $\sigma^\text{School}$. Likewise, sampling from the posterior of $\gamma_{j}^\text{School}$ for an unobserved school $j$ is done by sampling from a normal distribution with mean implied by the posteriors $\gamma_{j^*}^\text{School}$ of all the observed schools $j^*$, and standard deviation implied by the posterior distribution of $\sigma_\gamma^\text{School}$.

\subsection{Comparison metrics}

Point estimation quality, standard error quality, and quality of uncertainty interval is evaluated through the three metrics when estimating CATE for subpopulation $\mathcal{O}$. Let $M$ be the number of simulation iterations. We will sample $\mathcal{D}_1,\dots, \mathcal{D}_M$ from the same target population $P$ ($P$ resets after each sample) in each new simulation iteration based on the same stratified sampling scheme.

For each of the six estimation methods, based on sample $\mathcal{D}_m$, define $\mathbb{E}\left(\hat{\tau}_{\text{CATE},\mathcal{O}} | \mathcal{D}_m\right)$ to be the point estimate for $\tau_{\text{CATE},\mathcal{O}}$, and $\hat{\sigma}_{\text{m,CATE},\mathcal{O}}$ to be the standard error that accompanies the point estimate.

\subsubsection{Mean squared error for $\tau_{\text{CATE},\mathcal{O}}$ (MSE)}

MSE measures the quality of the point estimate for each of the 6 treatment effect estimation methods in our simulation study. For each method and for every simulation iteration, we calculate the empirical MSE in Equation \ref{eqn:mse}: 

\begin{equation}
\begin{aligned}
L_\text{MSE}(\mathcal{D}_1,\dots,\mathcal{D}_M)&:=\frac{1}{M}\sum_{m=1}^M \left( \mathbb{E}\left(\hat{\tau}_{\text{CATE},\mathcal{O}} | \mathcal{D}_m\right) - \tau_{\text{CATE},\mathcal{O}} \right)^2\\
&=\left( \frac{1}{M}\sum_{m=1}^M\mathbb{E}\left(\hat{\tau}_{\text{CATE},\mathcal{O}} | \mathcal{D}_m\right) - \tau_{\text{CATE},\mathcal{O}}\right)^2 \\
&+ \frac{1}{M}\sum_{m=1}^M \left(\mathbb{E}\left(\hat{\tau}_{\text{CATE},\mathcal{O}} | \mathcal{D}_m\right) - \frac{1}{M}\sum_{m'=1}^M\mathbb{E}\left(\hat{\tau}_{\text{CATE},\mathcal{O}} | \mathcal{D}_{m'}\right) \right)^2
\end{aligned}
\label{eqn:mse}
\end{equation}

\subsubsection{Proper scoring rule for $\tau_{\text{CATE},\mathcal{O}}$ (PSR)}

To simultaneously evaluate the quality of a point estimate and the accompanying standard error estimate, we utilize a metric from decision theory, which only depends on the first and the second moment of a probabilistic forecast. Consider the following metric \cite{gneiting2007strictly}:

\begin{equation}
    \begin{aligned}
    L_\text{PSR}(\mathcal{D}_1,\dots,\mathcal{D}_M)&:=-\frac{1}{M}\sum_{m=1}^M\left(\frac{\tau_{\text{CATE},\mathcal{O}}-\mathbb{E}\left(\hat{\tau}_{\text{CATE},\mathcal{O}} | \mathcal{D}_m\right) }{\hat{\sigma}_{\text{m,CATE},\mathcal{O}}}\right)^2 - \frac{1}{M}\sum_{m=1}^M\text{log}(\hat{\sigma}_{\text{m,CATE},\mathcal{O}}^2)
    \end{aligned}
\end{equation}

A higher PSR value corresponds to a higher quality prediction based off this metric.

\subsubsection{Coverage probabililty of uncertainty intervals}

Lastly, we will measure the quality of the uncertainty interval generated by the estimation method. Let $l_{0.025}^{(m)} \left(\hat{\tau}_{\text{CATE},\mathcal{O}} \right), u_{0.975}^{(m)} \left(\hat{\tau}_{\text{CATE},\mathcal{O}} \right)$ be the bounds of the symmetric 95-percent uncertainty interval for the point estimate $\mathbb{E} \left(\hat{\tau}_{\text{CATE},\mathcal{O}} | \mathcal{D}_m \right)$. The coverage probability of the uncertainty interval is then:

\begin{equation}
    \begin{aligned}
    L_\text{Coverage}(\mathcal{D}_1,\dots,\mathcal{D}_M)&:=\frac{1}{M}\sum_{m=1}^M \mathbbm{1}\left(l_{0.025}^{(m)} \left(\hat{\tau}_{\text{CATE},\mathcal{O}} \right) \le \tau_{\text{CATE},\mathcal{O}} \le u_{0.975}^{(m)} \left(\hat{\tau}_{\text{CATE},\mathcal{O}} \right)  \right)
    \end{aligned}
\end{equation}

\subsection{Estimating the ATE, CATE for highest achieving schools $\times$ lowest minority composition and CATE for African Americans in lowest achieving schools}

\begin{figure}[ht]
    \centering
    \includegraphics[width=1\textwidth]{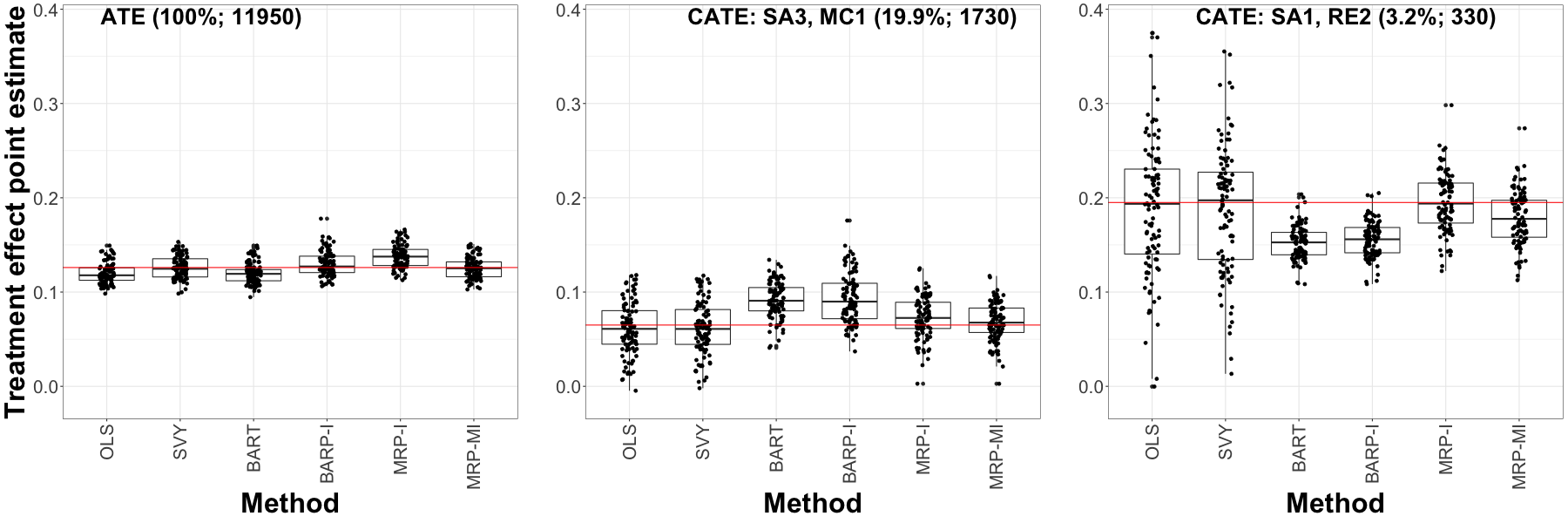}
    \caption{Left: Point estimate of ATE based on the full stratified cluster sample of around 12000 observations. Middle: Point estimate of CATE for individuals in the highest-achieving schools with the lowest minority composition. Right: Point estimate of CATE for individuals in the lowest-achieving schools and are African-Americans. The labels in the top-left of each plot are the proportion of the population that the group represents and the expected sample size of that group in $\mathcal{D}$. 100 simulations were conducted. The red horizontal line is the true treatment effect $\tau_{\text{CATE},\mathcal{O}}$ for the specific subpopulation $\mathcal{O} $ on the plot.}
    \label{fig:ate_2cates_plot}
\end{figure}

Figure \ref{fig:ate_2cates_plot} shows the 5 competing methods, along with MRP-MI and their point estimates for the ATE and two CATEs. As the target group in the population gets smaller, the traditional treatment effect estimation methods, SVY and OLS in particular, produce more variable point estimates. 

As the area gets smaller, to the point where we are looking at 3.2$\%$ of the population, MRP-MI's point estimates of the target CATE remain tight around the truth. The frequentist regression methods SVY and OLS exhibit higher spread in point estimates (though they remain unbiased). This isn't surprising, as Bayesian methods (MRP-MI, BARP-I, BART, MRP-I) tend to outperform frequentist approaches (SVY, OLS) in the presence of limited data. In contrast to this, calculating the ATE uses the full sample $\mathcal{D}$ and this corresponds to 100$\%$ of the population. All the methods except MRP-I have point estimate quality as nearly identical when calculating the ATE.

In Figure \ref{fig:ate_2cates_plot}, we visualized point estimates for the ATE and two CATEs of varying sizes in the target population (size is in reference to the percent of the population in the top-left of the three plots in Figure \ref{fig:ate_2cates_plot}). Table \ref{tab:ate_twocates_table} contains the MSE and PSR for each of the six methods. MRP-MI is shown to have the highest PSR based on Table \ref{tab:ate_twocates_table}, implying that it strikes the best balance of producing accurate point estimates with reasonable standard errors. 
Based on Table \ref{tab:ate_twocates_table}, MRP-MI is shown to have the best MSE, except for calculating the CATE corresponding to the group of individuals who are in the lowest achieving schools and are African-American (Race/Eth. coded as 2). However, the MSE of MRP-MI (10.70) is still close to being the lowest for this CATE (9.60). 

An important detail to note is that, even though MRP-I appears to be very similar to MRP-MI in point estimate quality when just looking at Figure \ref{fig:ate_2cates_plot}, we can see from Table \ref{tab:ate_twocates_table} that the MSE values for the ATE are drastically different. MRP-MI's MSE is more than twice as small as MRP-I's MSE when the target estimand is the ATE. This decrease in point estimation quality is attributed to MRP-I not accounting for the full Prev-GPA distribution in every poststratification cell $c$. It appears that this MSE difference between MRP-MI and MRP-I gets smaller as the target subpopulation becomes a smaller proportion of the target population. 
\begin{table}[ht]
\centering
\resizebox{\columnwidth}{!}{
\begin{tabular}{lrrrrrrr}
  \hline
Model & ATE (MSE $\times 10^{-4}$) & ATE (PSR) & SA3 MC1 (MSE $\times 10^{-4}$) & SA3 MC1 (PSR) & SA1 RE2 (MSE $\times 10^{-4}$) & SA1 RE2 (PSR) \\ 
  \hline
OLS & 1.59 & 7.30 & 7.83 & 6.12 & 50.44 & 4.31 \\ 
SVY & 1.40 & 7.83 & 7.94 & 5.87 & 46.29 & 4.33 \\ 
BART & 1.64 & 7.54 & 10.90 & 3.18 & 21.55 & 0.13 \\ 
BARP-I & 1.97 & 7.55 & 15.17 & 3.67 & 19.11 & 3.05 \\
MRP-I & 2.97 & 7.07 & 5.55 & 6.44 & \textbf{9.60} & 5.81 \\ 
MRP-MI & \textbf{1.28} & \textbf{7.94} & \textbf{4.35} & \textbf{6.71} & 10.70 & \textbf{5.83} \\ 
   \hline
\end{tabular}
}

\caption{Mean square error ($L_\text{MSE}$) and the proper scoring rule ($L_\text{PSR}$) value for the ATE, CATE for the group of individuals who are in the highest achieving schools with the lowest minority composition, and CATE for the group of individuals who are in the lowest achieving schools and are African-American. 100 simulations were conducted. The best metric value in each column is in bold. Values are rounded to two decimal places.}
\label{tab:ate_twocates_table}
\end{table}

The coverage probabilities in Table \ref{tab:ate_twocates_table_ci} shows that when calculating the ATE, all six methods have well-calibrated uncertainty intervals as they are all close to the nominal value of 95$\%$. As the target subpopulation gets smaller (size is in reference to the percent of the population in the top-left of the three plots in Figure \ref{fig:ate_2cates_plot}), we see that BARP-I and BART have poor calibration. The frequentist methods SVY and OLS retain intervals with 95$\%$ coverage probability.

\begin{table}[ht]
\centering
\begin{tabular}{lrrrr}
  \hline
Model & ATE & SA3 MC1 & SA1 RE2 \\ 
  \hline
OLS & 0.93 & 0.93 & 0.94 \\ 
SVY & 0.97 & 0.91 & 0.93 \\ 
BART & 0.90 & 0.58 & 0.45 \\ 
BARP-I & 0.91 & 0.76 & 0.60 \\
MRP-I & 0.87 & 0.98 & 0.99 \\ 
MRP-MI & 0.97 & 0.97 & 0.98 \\ 
   \hline
\end{tabular}
\caption{Coverage probabilities $L_\text{coverage}$ of 95 uncertainty intervals based on the various treatment effect estimation methods. The estimands are the ATE, CATE for the group of individuals who are in the highest achieving schools with the lowest minority composition, and CATE for the group of individuals who are in the lowest achieving schools and are African-American. 100 simulations were conducted.}
\label{tab:ate_twocates_table_ci}
\end{table}

In general, we see that MRP-MI is able to capture the true heterogeneity of treatment effects based on Figure \ref{fig:ate_2cates_plot} and the MSE columns of Table \ref{tab:ate_twocates_table}, even for areas as small as 3.2$\%$ of the target population. MRP-MI produces low variance and low absolute bias estimates as seen in Figure \ref{fig:ate_2cates_plot}. The coverage of the 95$\%$ uncertainty intervals for MRP-MI remains well-calibrated and hence close to 95$\%$ coverage when calculating the ATE (whole target population) and calculating CATEs for smaller subpopulations of the target population. MRP-MI appears to have the highest quality for point estimation, standard errors, and interval coverage.

The more simplistic methods MRP-I and BARP-I do not account for the full distribution of the continuous non-census variable Prev-GPA in every poststratification cell as they impute only one Prev-GPA value, whereas MRP-MI accounts for the full distribution of Prev-GPA at the cost of additional computation. The effect of not accounting for the full Prev-GPA distribution is most drastic when the estimand is the ATE. MSE, PSR and confidence interval coverage of BARP-I and MRP-I is worse than for MRP-MI. Studying the effects of not accounting for the full distribution of a non-census variable is a subject of future MRP research.

\subsection{Estimating the CATEs of the five subpopulation groups in highest achieving schools $\times$ lowest minority composition $\times$ race/ethnicity}

To further test the extent that MRP-MI and the other five methods capture treatment effect heterogeneity for small areas, we look at subpopulations as small as 1.3$\%$ of the target population.

\begin{figure}[ht]
    \centering
    \includegraphics[width=1\textwidth]{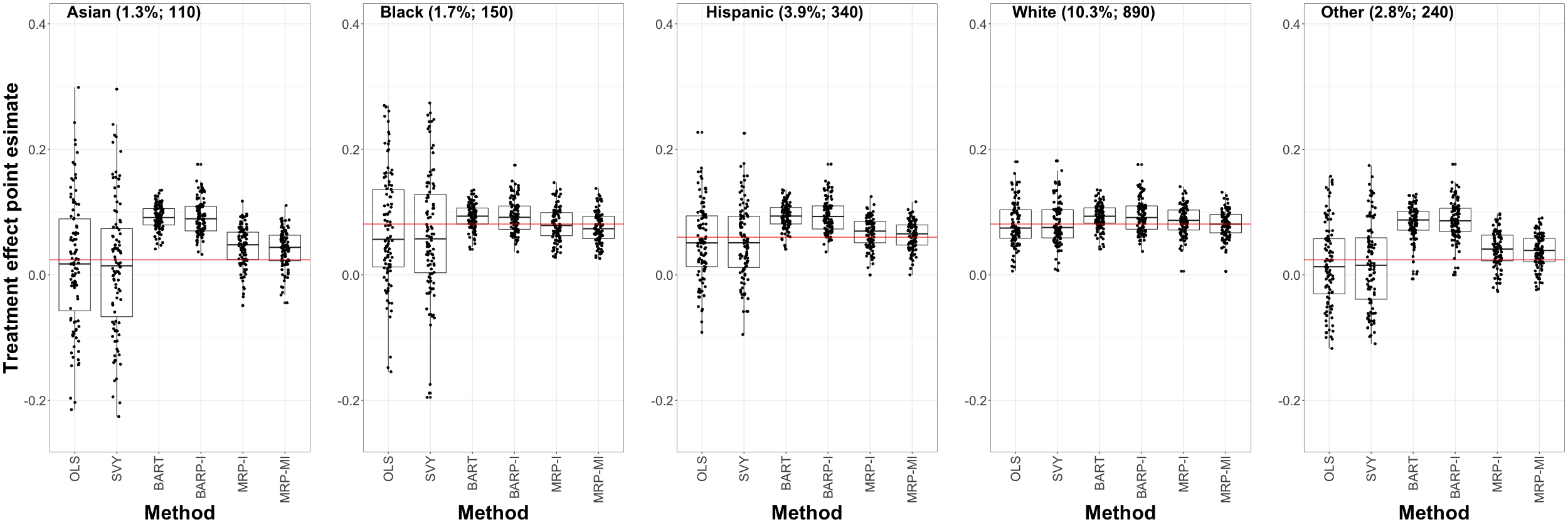}
    \caption{Point estimates of $\tau_{\text{CATE},\mathcal{O}}$ where $\mathcal{O}$ corresponds to the five race/ethnicities in the stratum highest achieving schools $\times$ lowest minority composition. 100 simulations were conducted. The labels in the top-left of each plot are the proportion of the population that the group
represents and the expected sample size for that group in $\mathcal{D}$. The red horizontal line is the true treatment effect $\tau_{\text{CATE},\mathcal{O}}$ for the specific subpopulation $\mathcal{O} $ on the plot.}
    \label{fig:sa3_mc1_raceeth}
\end{figure}

Specifically, the target subpopulations are the five race/ethnicities in the stratum highest achieving schools $\times$ lowest minority composition. Figure \ref{fig:sa3_mc1_raceeth} shows the CATE point estimates for CATEs of these five subpopulations.

Figure \ref{fig:sa3_mc1_raceeth} shows the five subpopulations ranging from 1.3$\%$ of the population to 10.3$\%$ of the population. Their expected sample sizes in $\mathcal{D}$ vary from 110 to 890 individuals. One average, the response rates are equal amongst the five subpopulations in the stratum of highest achieving schools with the lowest minority composition based off our experimental design for collecting $\mathcal{D}$.

In the stratum of highest achieving schools with the lowest minority composition, the subpopulation with the smallest expected sample size of 110 in $\mathcal{D}$ are Asian individuals. From Figure \ref{fig:sa3_mc1_raceeth} we see that the Bayesian estimation methods (MRP-MI, MRP-I, BART, BARP-I) have lower variance amongst their point estimates than the frequentist methods' (SVY, OLS) point estimates. 

In the stratum of highest achieving schools with the lowest minority composition, the subpopulation with the highest expected sample size of 890 in $\mathcal{D}$ are White individuals. From Figure \ref{fig:sa3_mc1_raceeth}, we see that the Bayesian methods (MRP-MI, MRP-I, BART, BARP-I) appear more similar to the frequentist methods (SVY, OLS). Unsurprisingly, the frequentist methods (SVY, OLS) have the highest variance estimates in the presence of limited data such as in the subpopulations of Asian individuals and African-American individuals in Figure \ref{fig:sa3_mc1_raceeth}. 

From Figure \ref{fig:sa3_mc1_raceeth} MRP-MI and MRP-I appear quite similar and both have a good balance of generating low-variance and low-bias estimates. This implies that the MSE of MRP-MI and MRP-I should be scored highly amongst the competing methods. Indeed this is true from Table $\ref{tab:sa3_mc1_raceeth}$ --- MRP-MI has the best MSE for all five CATEs and MRP-I has the second best MSE for all five CATEs. The shrinkage of the Bayesian nonparametric tree-based methods (BART, BARP-I) with their default priors appear to be too strong for estimating the heterogeneous treatment effects of small areas, since they are shown to be the most biased yet have the lowest variance in the five panels of Figure \ref{fig:sa3_mc1_raceeth}.

\begin{table}[ht]
\centering
\begin{tabular}{lrrrrrrrrrr}
  \hline
Model & RE1 & RE2 & RE3 & RE4 & RE5 \\ 
  \hline
OLS (MSE $\times 10^{-4}$) & 118.90 & 93.98 & 36.09 & 12.07 & 46.79 \\ 
SVY (MSE $\times 10^{-4}$) & 103.65 & 81.45 & 36.88 & 12.21 & 45.08 \\ 
BART (MSE $\times 10^{-4}$) & 54.83 & 9.12 & 18.79 & 8.99 & 46.32 \\ 
BARP-I (MSE $\times 10^{-4}$) & 54.13 & 8.76 & 19.38 & 8.95 & 49.49 \\
MRP-I (MSE $\times 10^{-4}$) & 14.27 & 7.10 & 6.29 & 5.56 & 10.70 \\ 
MRP-MI (MSE $\times 10^{-4}$) & \textbf{11.74} & \textbf{6.58} & \textbf{5.06} & \textbf{4.58}& \textbf{8.81} \\
  \hline

OLS (PSR) &  3.49 & 3.68 & 4.60  & 5.78 &  4.34 \\ 
SVY (PSR) & 3.57 & 3.78 & 4.21 & 5.61 & 3.91 \\ 
BART (PSR) & -9.05 & 5.94 & 1.46 & 5.88 & -7.08 \\ 
BARP-I (PSR) & -2.87 & 5.60 & 3.31 & 5.52 & -1.98 \\
MRP-I (PSR) & 5.46 & 6.12 & 6.25 & 6.41 &  5.80 \\ 
MRP-MI (PSR)  &  \textbf{5.64}  & \textbf{6.30}  & \textbf{6.52} & \textbf{6.66} & \textbf{5.96} \\
   \hline
\end{tabular}
\caption{Mean square error ($L_\text{MSE}$) and the average proper scoring rule value ($L_\text{PSR}$) based on the various treatment effect estimation methods. The first six rows correspond to $L_\text{MSE}$ estimates and the last six rows correspond to $L_\text{PSR}$ estimates. The estimands are CATEs for all five race/ethnicities in the subpopulation of highest achieving schools $\times$ lowest minority composition. RE1 corresponds to Asian individuals, RE2 corresponds to African-American individuals, RE3 corresponds to Hispanic individuals, RE4 corresponds to White individuals, RE5 corresponds to individuals of other race/ethnicity. 100 simulations were conducted. The best metric value in each column is in bold. Values are rounded to two decimal places}
\label{tab:sa3_mc1_raceeth}
\end{table}

Table \ref{tab:sa3_mc1_raceeth} further shows that MRP-MI has the best PSR values for each of the five CATEs. There is a large drop off in estimation quality for the Bayesian nonparametric tree-based methods (BART, BARP-I), as seen by their low PSR values for calculating CATEs of Asian individuals (RE1) and individuals of other race/ethnicity (RE5). 

Table \ref{tab:sa3_mc1_raceeth_ci} shows that MRP-MI has well-calibrated 95$\%$ uncertainty intervals despite the heterogeneity of treatment effects across the five small subpopulations. The Bayesian nonparametric tree-based methods (BARP-I, BART) have the worst calibration as their intervals' coverage probabilities are quite far from 95 percent. 
\begin{table}[ht]
\centering
\begin{tabular}{lrrrrrr}
  \hline
 Model & RE1 & RE2 & RE3 & RE4 & RE5 \\ 
  \hline
OLS & 0.94 & 0.94 & 0.96 & 0.98 & 0.99 \\ 
SVY & 0.95 & 0.92 & 0.89 & 0.94 & 0.93 \\ 
BART & 0.21 & 0.84 & 0.50 & 0.82 & 0.34 \\ 
BARP-I & 0.28 & 0.90 & 0.72 & 0.89 & 0.38 \\
MRP-I & 0.99 & 0.99 & 1.00 & 0.98 & 0.96 \\ 
MRP-MI & 0.99 & 0.98 & 0.98 & 0.97 & 0.94 \\ 
   \hline
\end{tabular}
\caption{Coverage probabilities $L_\text{coverage}$ of 95 uncertainty intervals based on the various treatment effect estimation methods. The estimands are CATEs for the five race/ethnicities in the subpopulation highest achieving schools $\times$ lowest minority composition. RE1 corresponds to Asian individuals, RE2 corresponds to African-American individuals, RE3 corresponds to Hispanic individuals, RE4 corresponds to White individuals, RE5 corresponds to individuals of other race/ethnicity. 100 simulations were conducted.}
\label{tab:sa3_mc1_raceeth_ci}
\end{table}

In summary, MRP-MI appears to be the most effective in estimating treatment effects for very large areas (ATE) to very small areas in the target population when there is treatment effect heterogeneity. The accompanying standard errors and uncertainty intervals produced by MRP-MI are also well-calibrated. Frequentist methods (SVY, OLS) appear to perform well when the observed sample $\mathcal{D}$ is sufficiently large but have extremely high variance estimates when they're used to calculate CATEs for small-areas, as data is more limited. Bayesian nonparametric tree-based methods (BART, BARP-I) have default priors that exhibit very strong shrinkage when compared to hierarchical modeling approaches (MRP-MI, MRP-I) for calculating heterogeneous CATEs in small areas and furthermore their uncertainty intervals have poor coverage when the target area gets smaller. A further area of research would be to perform prior calibration such as prior predictive checks for these Bayesian nonparametric methods.

Even though the point estimates Figures \ref{fig:ate_2cates_plot} and \ref{fig:sa3_mc1_raceeth} show that MRP-I and BARP-I have comparable point estimates to MRP-MI, we do not recommend using MRP-I and BARP-I for a causal inference problem similar to the one in our simulation study. They do not account for the full distribution of the continuous covariate Prev-GPA, whereas MRP-MI does.

MRP-MI, MRP-I and BARP-I get posterior predictive samples from $$\hat{\tau}_{\text{CATE},\mathcal{O}} =\frac{\sum_{c \in I_ \mathcal{O}} N_c \left( Y^{\text{rep},n}_{1,c} - Y^{\text{rep},n}_{0,c} \right)}{\sum_{c' \in I_ \mathcal{O}} N_{c'}}$$

As mentioned in Section 4, MRP-MI introduces a dependency structure (after conditioning on $\mathcal{D}$) to 

\begin{equation}
    \begin{aligned}
        \left(Y^{\text{rep},n}_{0,1},\dots,Y^{\text{rep},n}_{0,J},Y^{\text{rep},n}_{1,1},\dots,Y^{\text{rep},n}_{1,J} \right)
    \end{aligned}
    \label{yrep_vector}
\end{equation}

since the same $1000$ posterior samples from $\mathcal{M_\text{Prev-GPA}}$ and $\mathcal{M_\text{Post-GPA}}$ are used to sample from $Y^{\text{rep},n}_{z,c}$ for every $z$ and $c$. MRP-I and BARP-I use the same $1000$ posterior samples from their respective outcome model fits to get samples from $Y^{\text{rep},n}_{z,c}$ for every $z$ and $c$ as well, and hence a dependency structure exists in Equation \ref{yrep_vector} for MRP-I and BARP-I as well. The accompanying standard error $\sqrt{\mathbb{V} \left( \hat{\tau}_{\text{CATE},\mathcal{O}} | \mathcal{D} \right)}$ for the point estimate of $\tau_{\text{CATE},\mathcal{O}}$ is affected by the dependency structure in Equation \ref{yrep_vector} and studying this relationship for MRP-I, BARP-I and in particular MRP-MI, is an unexplored area.

\subsection{Limitations: Out of sample treatment effect estimation for the CATE of all 11221 schools in the population}

So far, the subpopulations $\mathcal{O}$ where CATEs are estimated are subpopulations that are observed in the stratified cluster sample $\mathcal{D}$. BARP-I, MRP-I and MRP-MI use the estimated distribution $\hat{\tau}_{\text{CATE},\mathcal{O}}$ to calculate CATE point estimates and this can be done even for $\mathcal{O}$ not observed in $\mathcal{D}$. In this subsection, we showcase the extent that MRP-MI is able to estimate $\tau_{\text{CATE},\mathcal{O}}$ for all the schools in the target population $P$.

\begin{figure}[ht]
    \centering
    \includegraphics[width=.6\textwidth]{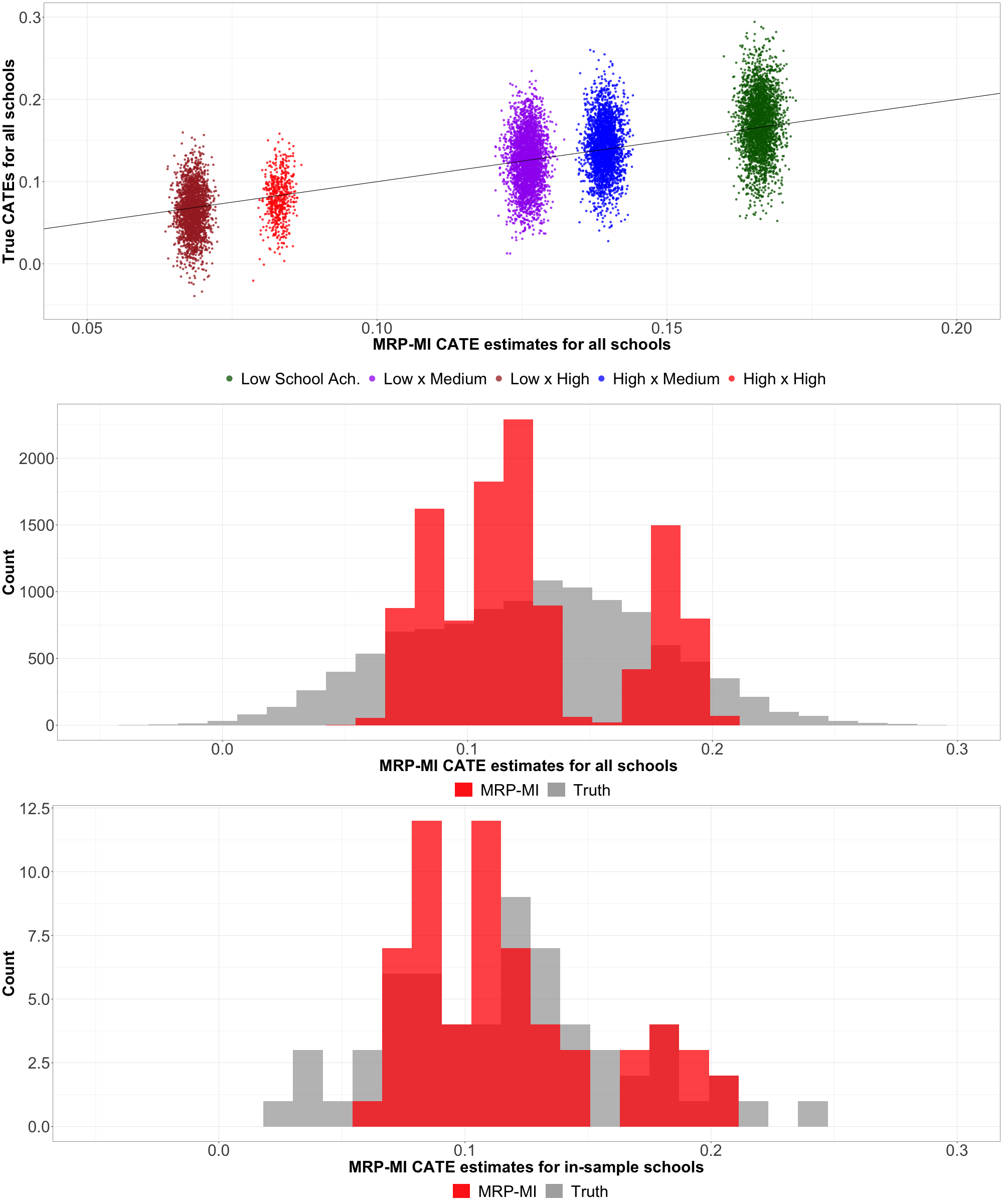}
    \caption{(Top) Average of CATE point estimates from MRP-MI for all 11221 schools based on 100 simulations. Every dot corresponds to a specific school. The x-axis are averages of our point estimates $\mathbb{E} \left( \hat{\tau}_{\text{CATE},\mathcal{O}} | \mathcal{D} \right)$. The y-axis are the true CATEs $\tau_{\text{CATE},\mathcal{O}}$ for every school. The dots are colour-coded by their school strata (Table \ref{tab:school_strata}).\
    (Bottom) The point estimates of the CATE for every school from MRP-MI for one simulation iteration.\ 
    (Middle) The point estimates of the CATE for all schools in $\mathcal{D}$ from MRP-MI, for the same simulation iteration as the middle histogram.
    }
    \label{fig:school_cates}
\end{figure}

Stratified cluster sample $\mathcal{D}$ in our simulation study contains around 70 schools for every new simulation iteration and thus there are around 11000 schools left unobserved. We are however given poststratification cell sizes $N_c$ for every cell $c$ in the population and thus are able to calculate $\hat{\tau}_{\text{CATE},\mathcal{O}}$ for every school $\mathcal{O}$ using MRP-MI (as well as MRP-I and BARP-I). Figure \ref{fig:school_cates} shows MRP-MI point estimates $\mathbb{E} \left(\hat{\tau}_{\text{CATE},\mathcal{O}} 
|\mathcal{D}\right)$ for every school in our simulation studies.

From the middle histogram of Figure \ref{fig:school_cates}, we see that point estimates of all 11221 schools $\mathbb{E} \left( \hat{\tau}_{\text{CATE},\mathcal{O}} | \mathcal{D} \right)$ from MRP-MI are not able to capture the full variation of CATEs $\tau_{\text{CATE},\mathcal{O}}$ using $\mathcal{D}$, even though we're given poststratification counts $N_c$ for all the cells $c$ in schools not in $\mathcal{D}$. In fact, the out-of-sample CATE estimates of MRP-MI in the middle histogram appear to be quite similar to the in-sample CATE estimates of MRP-MI in the bottom histogram. This implies that even with an external poststratification matrix on the full population, it remains difficult to calculate treatment effects for subpopulations not observed in the sample when using MRP-MI. We observed very similar challenges for MRP-I and BARP-I.

The top scatterplot further reinforces the claim that MRP-MI may be limited when calculating CATEs for schools not observed in $\mathcal{D}$. With the current amount of information provided through $\mathcal{D}$ and poststratification matrix $M$, MRP-MI (along with MRP-I and BARP-I) clusters CATE estimates for all the schools, clustering around the strata seen in Table \ref{tab:school_strata}. 

The clustering of CATE estimates seen in the top plot of Figure \ref{fig:school_cates} may be caused by how Prev-GPA was generated for $P$. In the data generating process for $P$, we had $V_{iks} \sim \mathcal{N}_{(0,4.33)}\left(\mu_X[s],\sigma_X[s]\right)$ for an individual $i$ in school $k$ of strata $s$. In fact, $V_{iks}$ is generated with only the knowledge of School Achievement level of individual $i$. This is seen from the DAG in Figure \ref{fig:dag}, since the only parent of the Prev-GPA node is School Achievement level. Hence the information on Prev-GPA in $\mathcal{D}$ may not be rich enough to capture the variability of CATEs for unobserved schools. Perhaps generating $V_{iks}$ for everyone in $P$ with additional variables will improve out of sample treatment effect estimation with MRP-style estimators based on $\mathcal{D}$, if $n$ is not increased. This would require adding more parent nodes to Prev-GPA in Figure \ref{fig:dag}.

\section{Conclusion}

The primary goal of this manuscript was investigating the capabilities of MRP-style estimators in experimental causal inference. Through simulation studies, we studied the limitations of MRP-style estimators in the problem of estimating heterogeneous treatment effects for subpopulations as small as 1.3 percent of the target population. We modeled our simulation studies after the National Student of Learning Mindsets study \cite{yeager2019national}, an experimental dataset that was fairly representative of target population, yet exhibited treatment effect heterogeneity.

We also build upon a variant of MRP (MRP-MI) that is able to incorporate non-census variables (along with their full distributions if they're continuous) into the poststratification step. In our simulations that modeled the NSLM study, MRP-MI along with other MRP-style estimators (MRP-I and BARP-I) are compared against modern Bayesian nonparametric causal inference methods \cite{hill2011bayesian} and frequentist regression methods \cite{lumley2020}. 

We showed that MRP-style estimators are effective at point estimation of conditional average treatment effects (CATE) for subpopulations of various sizes in the target population. When compared to the other treatment effect estimation methods in our simulations, MRP-MI strikes the best balance of having high-quality point estimates (low bias and variance in point estimates) and reasonable standard errors, along with well-calibrated uncertainty intervals for CATE estimation. The applications of the MRP-MI framework extend beyond causal inference and can be used in a traditional survey estimation problem where the modeler is solely interested in estimating population-level quantities, but this avenue of research is not considered in our manuscript.

Though the main focus of this manuscript was on treatment effect heterogeneity in small areas for stratified cluster experiments that are representative of the target population, we believe that MRP-style estimators such as MRP-MI can excel in dealing with nonrepresentative experimental data as well. MRP has traditionally been used as a model-based approach to generalizing survey estimates to a target population and we believe that this can be extended to generalization studies in experimentation. Comparing MRP-style estimators to propensity-score based generalization techniques in causal inference \cite{ackerman2020generalizing, ackerman2019implementing, stuart2018generalizability} is an open area of research. Extending the analysis done in this manuscript to the setting of observational causal inference is also an open area of research. 

When provided with a poststratification matrix of the target population along with experimental data, we showed through simulations that MRP-style estimators improve upon standard causal inference methods, especially in scenarios when the modeler wants to uncover treatment effect heterogeneity across small areas in the target population. MRP-MI is an effective framework for incorporating continuous non-census variables into poststratification. As such, we see MRP-MI as a useful approach for modelers when dealing with experimental data and poststratification information on not necessarily all of the variables in the target population.

\bibliographystyle{apalike}
\bibliography{biblio}

\newpage

\section*{Supplementary Material}

The GitHub repository for this manuscript is: \url{https://github.com/alexgao09/causal_mrp_public}. This repository will reproduce the results and plots in this manuscript.

\subsection*{A Proof of Proposition \ref{prop:poststratprobmeasure}}

\begin{proof}
\begin{equation}
    \begin{aligned}
        \mathbb{E}\left(\hat{\tau}_{\text{CATE},\mathcal{O}} |\mathcal{D} \right) & = \frac{\sum_{c \in I_\mathcal{O}} N_c \mathbb{E}\left( Y_{1,c}^{\text{rep},n} |\mathcal{D} \right)  - N_c \mathbb{E}\left( Y_{0,c}^{\text{rep},n} |\mathcal{D} \right) }{\sum_{c' \in I_\mathcal{O}} N_{c'}}\\
        &=\frac{\sum_{c \in I_\mathcal{O}} N_c \left( \int y^* p(y^* |1,c,\mathcal{D}) dy^* - \int y^* p(y^* |0,c,\mathcal{D}) dy^* \right) }{\sum_{c' \in I_ \mathcal{O}} N_{c'}}\\
        &=\frac{\sum_{c \in I_ \mathcal{O}} N_c \left( \int \int y^* p(y^* |1,c,v^*,\mathcal{D}) p(v^* | 1,c,\mathcal{D}) dy^* dv^* - \int \int y^* p(y^* |0,c,v^*,\mathcal{D}) p(v^* | 0,c,\mathcal{D}) dy^* dv^* \right) }{\sum_{c' \in I_ \mathcal{O}} N_{c'}}\\
        &=\frac{\sum_{c \in I_ \mathcal{O}} N_c \left( \int 
        \left( \int y^* p(y^* |1,c,v^*,\mathcal{D})dy^* - \int y^* p(y^* |0,c,v^*,\mathcal{D}) dy^* \right) p(v^* |c,\mathcal{D}) dv^* \right) }{\sum_{c' \in I_ \mathcal{O}} N_{c'}}\\
        &=\sum_{c \in I_ \mathcal{O}} \frac{N_c}{\sum_{c' \in I_ \mathcal{O}} N_{c'}} \int \left( \int y^* p(y^* |1,c,v^*,\mathcal{D})dy^* - \int y^* p(y^* |0,c,v^*,\mathcal{D}) dy^* \right) p(v^* |c,\mathcal{D}) dv^* \\
        &=\sum_{c \in I_ \mathcal{O}} \frac{N_c}{\sum_{c' \in I_ \mathcal{O}} N_{c'}} \int \left( g_1(c,v^*) - g_0(c,v^*) \right) p(v^* |c,\mathcal{D}) dv^*\\
    \end{aligned}
    \end{equation}
Defining $d \mu(c,v^*) := \frac{N_c}{\sum_{c' \in I_ \mathcal{O} } N_{c'}} p(v^* |c,\mathcal{D}) dv^*$ where the support is $(c,v^*) \in I_\mathcal{O} \times (0,4.33)$, we see that $\mathbb{E}\left(\hat{\tau}_{\text{CATE},\mathcal{O}} |\mathcal{D} \right) = \int \left( g_1(c,v^*) - g_0(c,v^*) \right) d \mu(c,x^*) $.

\end{proof}

\subsection*{B Further discussion on connection between MRP and treatment effect estimation}

Suppose that we have an estimation method $\mathcal{M}$ that produces unbiased estimates $\mathbb{\hat{E}}\left(Y(1)-Y(0) | X = x_c \right)$ for the true treatment effect $\tau_{\text{CATE},X=x}$ in every poststratification cell $c$. 

Let $\text{NR}\left( \mathcal{O} \right) \subseteq I_\mathcal{O}$ be the set of poststratification cells such that the in-sample frequency $\frac{n_{c^*}}{\sum_{c' \in I_\mathcal{O}} n_{c'}}$ differs from the true frequency $\mathbb{P}\left(X = x_{c^*} | X \in \mathcal{O} \right) = \frac{N_{c^*}}{\sum_{c' \in I_\mathcal{O}} N_{c'}}$. This means that the observed sample of covariates do not generate the true frequencies observed in the finite target population for cells in $\text{NR}\left( \mathcal{O} \right)$.

From Equation \ref{eqn:cateOinsample}, the point estimate of $\tau_{\text{CATE},\mathcal{O}}$ minus the truth is

\begin{equation}
    \begin{aligned}
        \sum_{c \in I_\mathcal{O}} \underbrace{\mathbb{\hat{E}}\left(Y(1)-Y(0) | X = x_c \right)}_\text{Point estimate from $\mathcal{M}$} \left( \frac{n_c}{\sum_{c' \in I_\mathcal{O}}n_{c'}} \right) -\tau_{\text{CATE},\mathcal{O}} & = \sum_{c \in I_\mathcal{O}} \tau_{\text{CATE},X=x} \left( \frac{n_c}{\sum_{c' \in I_\mathcal{O}}n_{c'}} \right) \\ 
        &-\sum_{c \in I_\mathcal{O}} \tau_{\text{CATE},X=x} \left( \frac{N_c}{\sum_{c' \in I_\mathcal{O}}N_{c'}} \right)\\
        &= \sum_{c \in \text{NR} \left(\mathcal{O} \right)} \tau_{\text{CATE},X=x} \left( \frac{n_c}{\sum_{c' \in I_\mathcal{O}}n_{c'}} - \frac{N_c}{\sum_{c' \in I_\mathcal{O}}N_{c'}} \right)
    \end{aligned}
    \label{eqn:cateOinsample_expanded}
\end{equation}

Hence the bias for estimating $\tau_{\text{CATE},\mathcal{O}}$ is driven by the size and direction of the true treatment effects $\tau_{\text{CATE},X=x_{c^*}}$ and the differences in the proportions $\frac{n_{c^*}}{\sum_{c' \in I_\mathcal{O}}n_{c'}} - \frac{N_{c^*}}{\sum_{c' \in I_\mathcal{O}}N_{c'}}$ for all ${c^*} \in \text{NR} \left( \mathcal{O} \right)$. The differences in proportions is caused by the observed sample of covariate vectors $\{ X_i\}_{i=1}^n$ being a nonrepresentative sample of individuals in the target population, $\{ X_i\}_{i=1}^{N_P}$.

\subsection*{C Coefficients for the data generating process in Section 5}

The coefficient vectors used in our simulation study outlined in Section 5 are defined as:

\begin{itemize}
    \item $\mu_X = \left(2.1, 2.8, 3.5\right)$. The three values correspond to the three school achievement levels, from low to high respectively. This comes from the sample mean of pre-intervention Core GPA seen on pages 31 - 33 of the appendix in \cite{yeager2019national}
    
    \item $\sigma_X = \left(0.8, 1, 0.6\right)$. The three values correspond to the three school achievement levels, from low to high respectively. This comes from the standard error of pre-intervention Core GPA seen on pages 31 - 33 of the appendix in \cite{yeager2019national}
    
    \item $p_\text{RE} = (0.026, 0.127, 0.280, 0.367, 0.200)$ for school stratum 1, 
    
    $p_\text{RE} = (0.053, 0.102, 0.234, 0.445, 0.166)$ for school stratum 2, 
    
    $p_\text{RE} = (0.023, 0.122, 0.254, 0.415, 0.186)$ for school stratum 3, 
    
    $p_\text{RE} = c(0.066, 0.086, 0.195, 0.514, 0.139)$ for school stratum 4,
    
    $p_\text{RE} = c(0.036, 0.106, 0.215, 0.484, 0.159)$ for school stratum 5.
    
    These vectors come from the empirical proportions of the five race/ethnicities shown on pages 31 - 33 of the appendix in \cite{yeager2019national}. (Asian, Black, Hispanic, White, Other)
    
    \item $p_\text{ME} = (0.238, 0.309, 0.386)$ for the lowest minority composition, $p_\text{ME} = (0.198 ,0.269, 0.346)$ for the highest minority composition. The three values in each vector correspond to the three school achievement levels, from low to high respectively. These vectors come from the empirical means of the Maternal College covariate shown on pages 31 - 33 of the appendix in \cite{yeager2019national}
    
    \item $\tau_{\text{SA}} = (0.1, 0.07, 0.01)$. The three values correspond to the three school achievement levels, from low to high respectively.
    
    \item $\tau_{\text{MC}} = (0, -0.01, 0.01)$. The two values correspond to the low and high minority composition levels respectively.
    
    \item $\tau_{\text{G}} = (0.01, 0.01)$. The two values correspond to the individual being Male or Female, respectively.
    
    \item $\tau_{\text{RE}} = (0.02, 0.1, 0.07, 0.1, 0.02)$. The five values correspond to the five race/ethnicities Asian, Black, Hispanic, White, Other respectively.
    
    \item $\tau_{\text{ME}} = (0.01, 0)$. The two values correspond to the individual having maternal education being True or False respectively.
\end{itemize}

$\tau_{\text{SA}}, \tau_{\text{MC}}, \tau_{\text{G}}, \tau_{\text{RE}}, \tau_{\text{ME}}$ were chosen so that the simulation setup in this manuscript produced treatment effects in the same range as the ones reported in \cite{yeager2019national}.

\end{document}